\documentclass[epj,twocolumn]{webofc}
\usepackage[varg]{txfonts}   
\woctitle{Jet dynamics: recollimation shocks and helical patterns.}
\begin{document}
\title{Jet dynamics and stability}
%
%
\subtitle{Recollimation shocks and helical patterns.}

\author{M. Perucho\inst{1}\fnsep\thanks{\email{manel.perucho@uv.es} }}
\institute{Departament d'Astronomia i Astrof\'{\i}sica, Universitat de Val\`encia}

\abstract{%
The dynamics and stability of extragalactic jets may be strongly influenced by small (and probable) 
differences in pressure between the jet and the ambient and within the jet itself. The former give rise to
expansion and recollimation of the jet. This occurs in the form of conical shocks, or Mach disks, if the 
pressure difference is large enough. Pressure asymmetries within the jet may trigger the development of helical patterns via coupling
to kink current-driven instability, or to helical Kelvin-Helmholtz instability, depending on the physical conditions in the jet. 
I summarize here the evidence collected during the last years on the presence of recollimation shocks 
and waves in jets. In the jet of CTA~102 evidence has been found for (traveling)shock-(standing)shock interaction in the core-region (0.1~mas from the core), using information from the light-curve of the source combined with VLBI data. The conclusions derived have been confirmed by numerical simulations combined with emission calculations that have allowed to study the spectral evolution of the perturbed jet. Helical structures can also be identified in radio-jets. The ridge-line of emission of the jet of S5~0836+710 has been identified as a physical structure corresponding to a wave developing in the jet flow. I review here the evidence that has allowed to reach this conclusion, along with an associated caveat. Current data do not allow to distinguish between magnetic or hydrodynamical instabilities. I finally discuss the importance of these linear and non-linear waves for jet evolution.}
\maketitle
\section{Introduction}
\label{intro}

  The parsec-scale structure of jets in active galactic nuclei (AGN) is mainly observed in the radio band, using the 
VLBI technique. These radio jets can be interpreted as flows because the Larmor radius of particles is very small 
when compared to the spatial scales of the problem studied \cite{br74}. However, it is very difficult to obtain information on 
the nature and properties of the underlying flow as related to the emitting region. Nevertheless, the interpretation of extragalactic jets as flows has different implications: Such systems, which are composed of magnetic fields and particles propagating along the jet channel are suitable for the growth of different hydrodynamical and/or magnetic instabilities. The instabilities have been claimed to be the cause of many of the structures observed in jets: knots produced by pinching, helices by kink or Kelvin-Helmholtz helical modes; see e.g., references \cite{ha06,ha11,pe12}. Those instabilities develop in the form of waves triggered by any external or internal perturbation that grow in amplitude with time and distance, as they are advected with the flow. The waves appear as solutions to the linearized equations of relativistic magneto-hydrodynamics \cite{bi91,pt05,ha07} (and references therein), giving solutions at most relevant wavelengths compared to the jet scale (0.1-100 jet radii). This means that any perturbation with a given frequency will almost certainly trigger the growth of an unstable mode and its harmonics \cite{pe05}. 

   The fact that FRII (Fanaroff-Riley type II, \cite{fr74}) jets propagate collimated to long distances poses a question on the effect 
of the instabilities on those jets. We now know that jet opening with distance or the 
growth of short-wavelength modes cancel the disruption of the jet via entrainment and deceleration, which would occur in the
case of full development of a relatively long wavelength mode to non-linear amplitudes, at least in the case of particle dominated jets \cite{ha06,ha11,pe12}. 
But, if instabilities do really grow in jets we would expect to observe wave-like structures and motions already with the available observing resolution provided by the interferometric technique. 

   One of the visible structures that could be generated in jets by the development of instabilities is the helical one. Actually, this kind of structure has been repeatedly reported in different jets (e.g., \cite{cm93,lz01}). Recently, \cite{pe+12} have proposed an alternative explanation for the helical structures observed in jets. Following previous work \cite{ha00,ha03}, the authors suggested that the jet flow may be forced into this path by the growth in amplitude of a perturbation that couple to an instability.  
 
   
  Considering the supersonic nature of jets, small pressure differences between the jet and the ambient generate the development of conical shocks produced by jet overexpansion with respect to the ambient \cite{dm88,fa91,ns09}. At these conical shocks, the jet flow may dissipate a fraction of its kinetic energy into thermal energy and emission \cite{go97,ko97,mi09,mo12}. If a Mach disk is formed (when the overpressure ratio and consequently the jet opening angle are large), the jet flow can even become subsonic and thus more unstable \cite{pm07} and eventually disrupted. Thus, in the case of parsec-scale jets I will focus on conical shocks to be consistent with the observed propagation of collimated jets beyond this region. The interaction of traveling perturbations with these standing shocks has been appointed as a possible scenario for the production of gamma-ray flares in AGN jets \cite{ma10,ag11} and it was already studied as a possible location for increase in radio flux \cite{go97,ko97}. 

  In the previous paragraphs I have set the theoretical justification for the presence of waves and standing shocks in jets. However, we were missing direct evidence of their real existence. Up to now, all the models were based on the assumption that observed stationary features could be identified with recollimation shocks, and on the assumption that instabilities must develop in jets to interpret helical structures as due to them. In the last years, evidence has been collected
for the presence of both in the jets in the blazar CTA~102 and the quasar S5~0836+710. On the one hand, C. Fromm and collaborators have shown that 
a flare in CTA~102 could have produced a perturbation traveling along the jet and interacting with a stationary feature within the core region (r$\simeq$0.1~mas) \cite{fr11,fr13,fr13a,fr13b}, via a combination of theoretical and observational analysis, together with numerical experiments. Recent emission simulations based on relativistic hydrodynamic simulations of an overpressured jet seem to confirm this interpretation (Fromm et al., in preparation) by studying the spectral evolution of the perturbed jet. On the other hand, observations of the jet in S5~0836+710 at different frequencies and epochs have shown that the ridge line of this jet behaves as expected if it is interpreted as a pressure wave \cite{pe+12}.

  Here I will summarize these results, discussing the implications for jet evolution and emission at parsec scales. I will also review them critically, pointing out the limitations of current studies of parsec-scale jets. The structure of this contribution is as follows: In Section~\ref{sec-1}, the accumulated evidence for shock-shock interaction in the jet in the blazar CTA~102 is exposed and the radiative signatures that this process could trigger is discussed. In Section~\ref{sec-2}, the evidence for the presence of waves in the jet in the quasar S5~0836+710 is reviewed. Finally, summary and a discussion on the implications that these structures may have for jet evolution are given in Section~\ref{sec-3}.


\section{Recollimation shocks}
\label{sec-1}

\subsection{Spectral evolution of a flare in CTA~102}
\label{ssec-1-1}
\begin{figure}
\centering
\includegraphics[width=\columnwidth,clip]{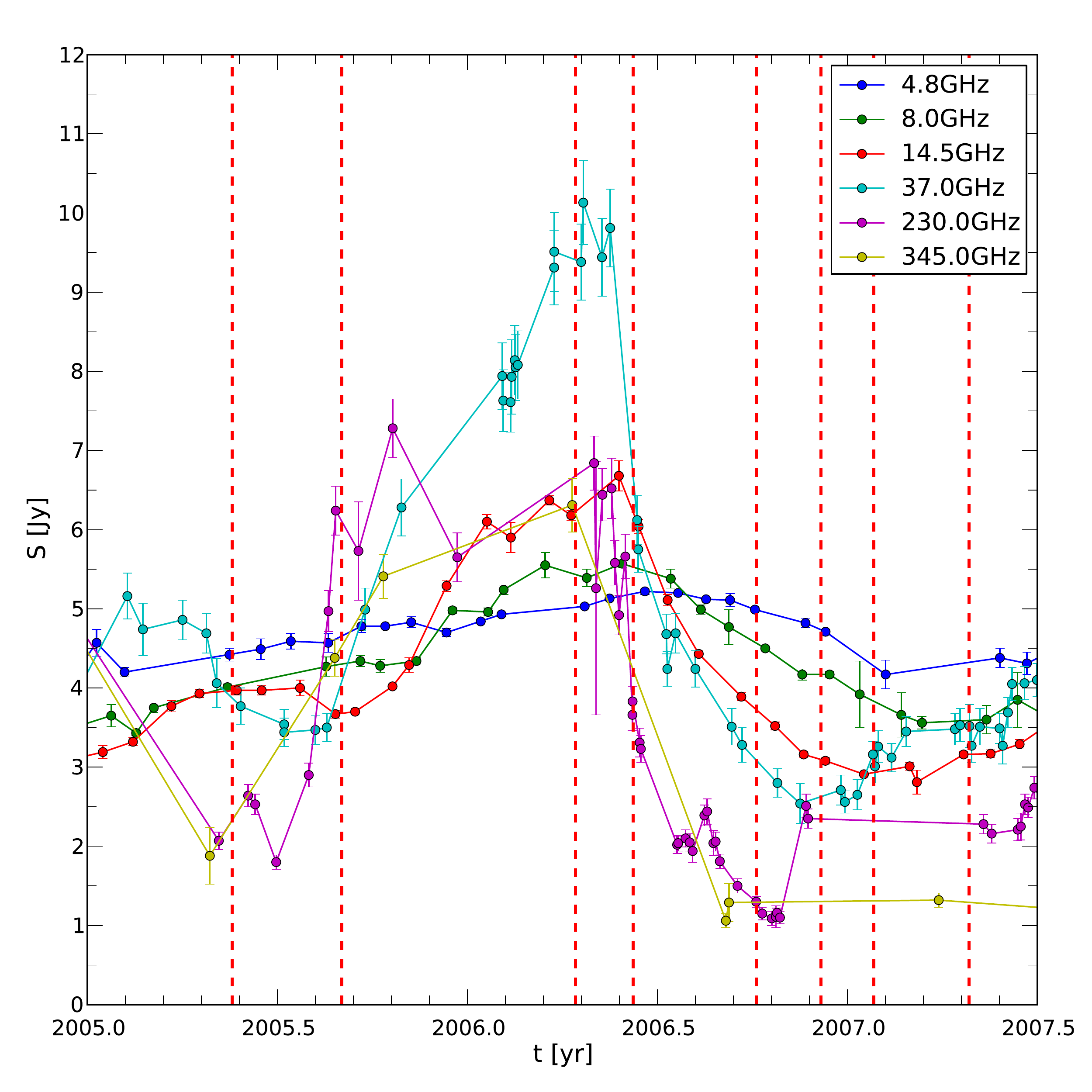}
\caption{Light curve of the jet in CTA~102 during the flare that took place in 2005/2006. The dashed lines indicate epochs of multi-wavelength VLBI observations. Taken from \cite{fr13b}.}
\label{fig-1}       
\end{figure}

 The light-curve of the jet in the blazar CTA~102 ($z=1.037$, $8.11$~pc/mas) showed an increase of almost an order of magnitude of the flux at radio frequencies, starting at the end of 2005 (see Fig.~\ref{fig-1}). A careful analysis of the properties of this flare can be found in \cite{fr11}. The authors identified an unexpected behavior in  the evolution of the turnover frequency and flux of this flare, if the normal shock-in-jet model evolution \cite{mg85} is assumed. This model predicts an initial increase of the turnover flux accompanied by a decrease of the turnover frequency during the initial stage of evolution, in which the losses are dominated by the inverse (self-)Compton (SSC) mechanism. As the flaring region evolves along the jet, it becomes more dilute and the inverse Compton scattering ceases to be the most important energy-loss mechanism, the turnover flux reaches its maximum and the turnover frequency continues decreasing. Finally, both turnover flux and frequency decrease in the last stage, in which the energy losses are mainly due to adiabatic expansion. In theory, there is a stage in which synchrotron losses are dominant between the Compton and the adiabatic stages, in which the turnover flux remains fairly constant, while the turnover frequency decreases. During the last two stages the spectral turnover is determined by synchrotron self-absorption (SSA). 
 
\begin{figure}
\centering
\includegraphics[width=\columnwidth,clip]{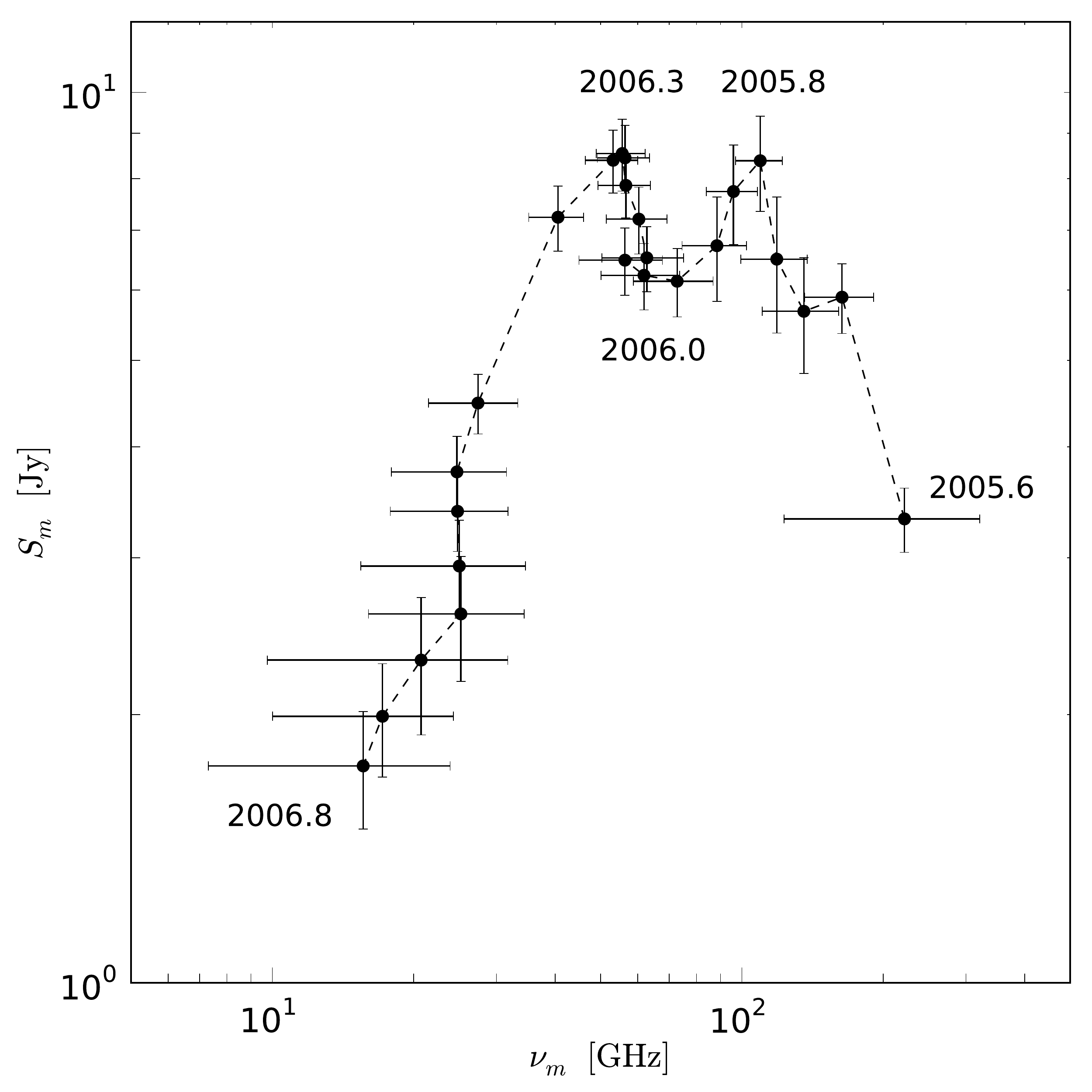}
\caption{Evolution of the turnover frequency-turnover flux of the flaring spectrum, which was obtained after subtracting the \emph{steady-state} fluxes, as defined in \cite{fr11}. Taken from \cite{fr11}.}
\label{fig-2}       
\end{figure}

  Within this picture, there should be a single maximum (or a plateau if there is a synchrotron stage) in the turnover flux density along the temporal evolution of the flare. However, in the case of CTA~102, the authors identified a double-peak structure, and no plateau in the evolution of the turnover frequency and flux, as shown in Fig.~\ref{fig-2}. The initial increase in flux was followed by a decrease, a new increase, and a second decrease in flux, with the turnover frequency always decreasing. Although during the second increase in flux it remained constant within errors. Once a change in the viewing angle or new flare were disregarded from modeling and observations, the hypothesis used to explain the second peak in flux was based on the possible interaction between the perturbation producing the flare and a stationary (recollimation) shock. The authors claimed that the initial Compton stage was followed by an adiabatic stage, which was interrupted by a second Compton stage. The interaction could produce reacceleration of particles and a second Compton stage, with an increase in the flux. Later, within the same proposed physical scenario, an adiabatic compression of the flow was claimed to possibly produce such increase in the flux \cite{fr13a,fr13b}. The reason being that the Compton mechanism requires larger opacity and particle density than probably available at this downstream region from the core. Nevertheless, the possibility of a second Compton stage cannot be completely ruled out.

\subsection{Evidence for shock-shock interaction}
\label{ssec-1-2}

\begin{figure*}
\centering
\includegraphics[width=\textwidth,clip]{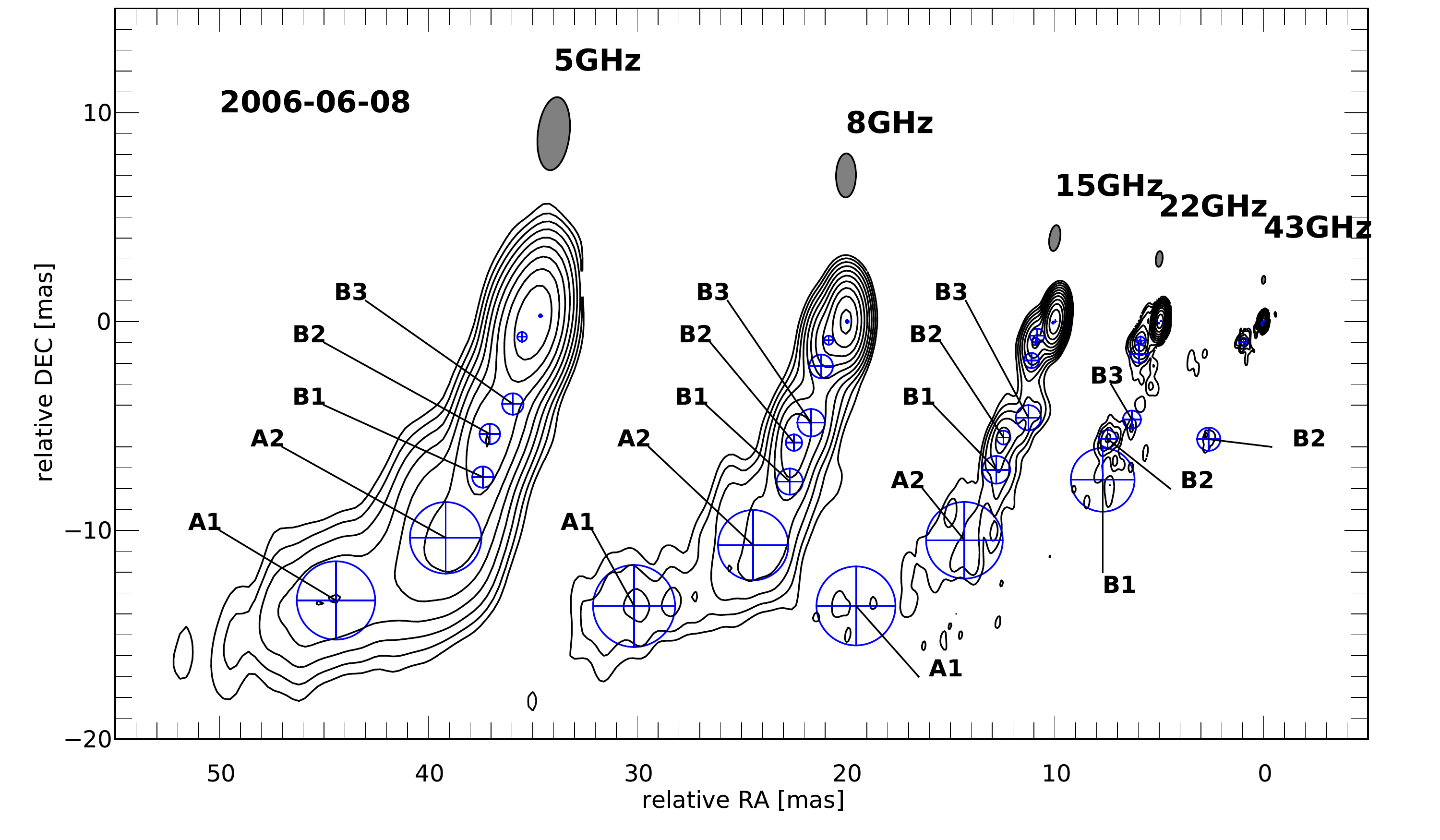}
\caption{The jet in CTA~102 at different frequencies and cross-identified components among frequencies. Taken from \cite{fr13a,fr13b}}
\label{fig-30}       
\end{figure*}

\begin{figure}
\centering
\includegraphics[width=\columnwidth,clip]{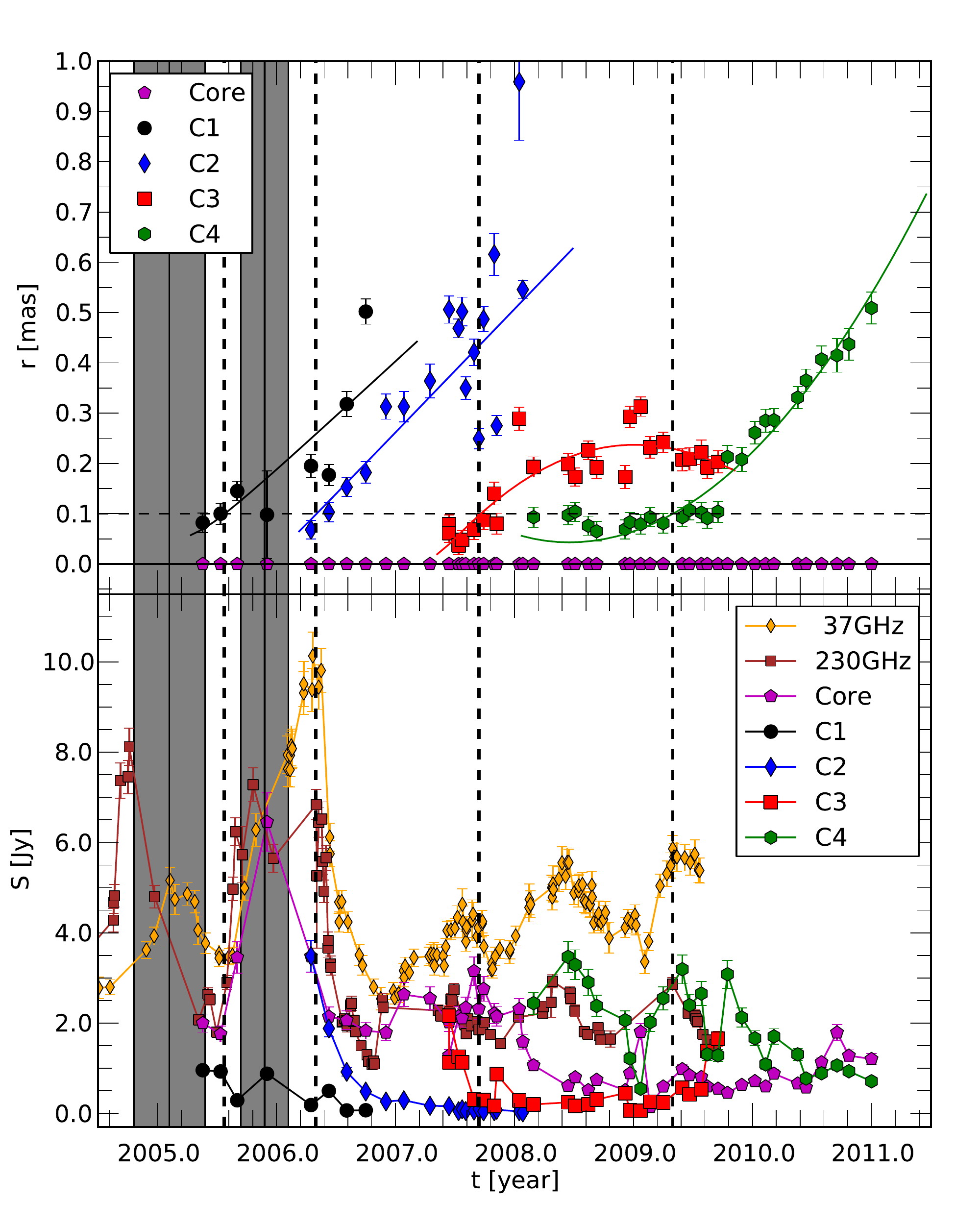}
\caption{Top panel: Kinematics of fitted components in the jet of CTA~102 at 43~GHz. The colors indicate the different identified components. Bottom panel: Simultaneous evolution of the light curve at different frequencies and of each of the fitted components. The shaded areas are identified with the injection of new components and the dashed lines with their passage through the region where a standing shock is expected. The component that is assumed to produce the studied flare is C2. Taken from \cite{fr13a}.}
\label{fig-3}       
\end{figure}

\begin{figure}
\centering
\includegraphics[width=\columnwidth,clip]{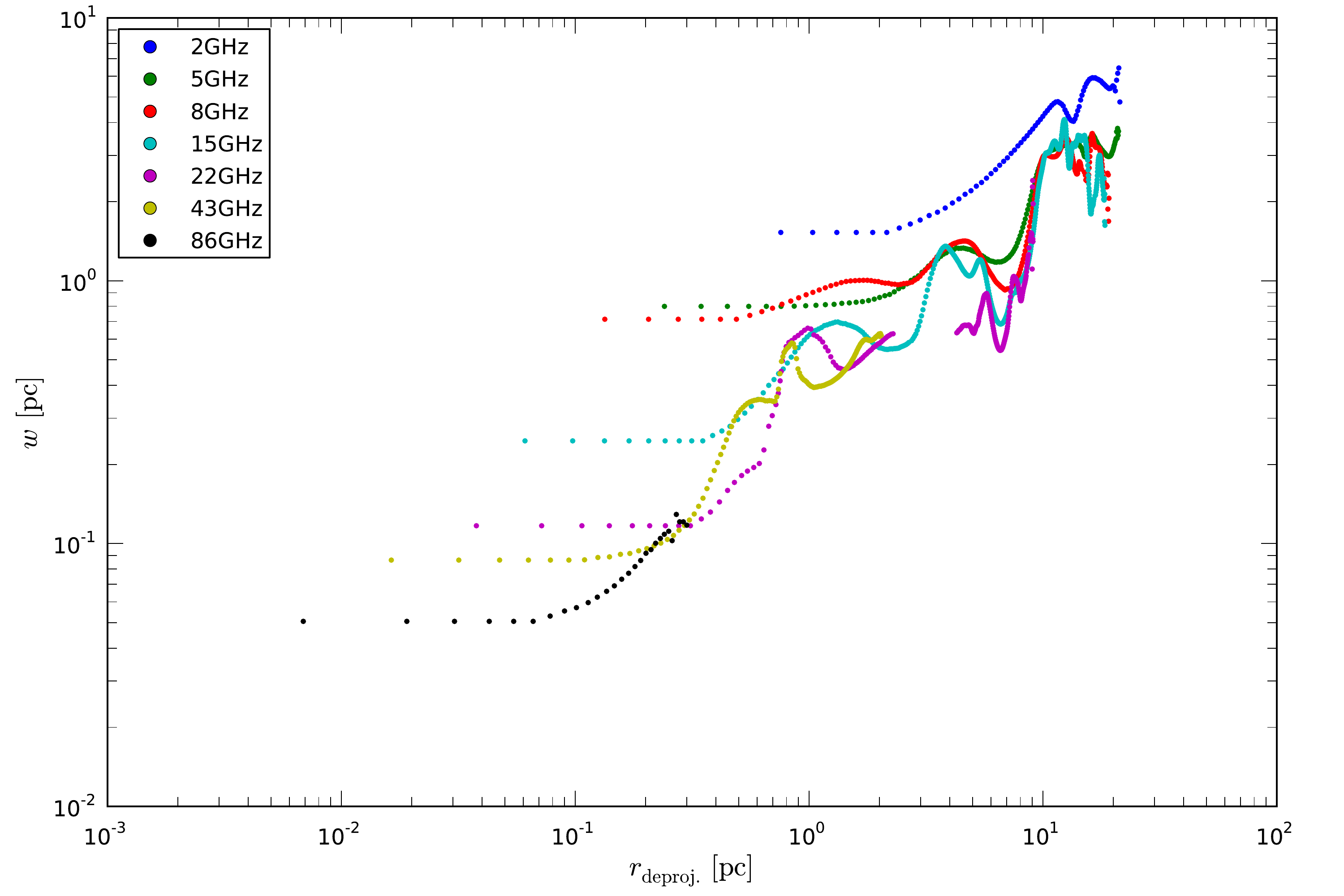}
\caption{Deconvolved jet radius with distance at different frequencies. Taken from \cite{fr13b}.}
\label{fig-4}       
\end{figure}

  Direct and indirect observational evidence for a recollimation shock in a region close to the radio core came from the multi-frequency analysis of the kinematics \cite{fr13a} and spectral behavior \cite{fr13b} of the source using VLBI observations (see Fig.~\ref{fig-30}). The summary of the collected evidence is the following: 1) An observable feature can be permanently fit in the analysis of the VLBI images at  43~GHz in a region around $0.1$~mas from the core (Fig.~\ref{fig-3}, \cite{fr13a}), even though it is typically identified with traveling features (but this could be wrong \cite{fr13}, Fromm et al. in preparation). 2) A traveling bright feature (component), injected at the beginning of the flare (late 2005), crosses this region during the time of the second peak in the turnover frequency-flux plane (component C2, Fig.~\ref{fig-3}, \cite{fr13a}). 3) At larger distances from the core, where the jet can be resolved, the jet radius shows successive expansions and compressions, which are a signature of non-pressure equilibrium with the ambient and thus, of the formation of recollimation shocks. Extrapolating this result inwards, we can expect to have such shocks within the core region (0-0.5~mas, Fig.~\ref{fig-4}, \cite{fr13b}). 4) A region in which the jet radius decreases and the resolution is enough to perform an accurate spectral analysis shows the expected spectral behavior of a recollimation shock, as obtained from numerical simulations, with a quasi-symmetric bumps in the turnover frequency and spectral index (Fig.~17 in \cite{fr13b}, \cite{mi09}). 5) The passage through this region of a previously injected feature was captured by the VLBI observations, showing a temporary increase in the observed turnover flux and frequency, too (Fig.~20 in \cite{fr13b}, \cite{fr13}). In conclusion, there is compelling evidence of the presence of recollimation shocks and common interaction with traveling bright features, which can be interpreted as shocks, in the jet of CTA~102, and hints of such an event at $r\simeq0.1$~mas from the radio-core. 
   
\begin{figure*}
\centering
\includegraphics[width=\textwidth,clip]{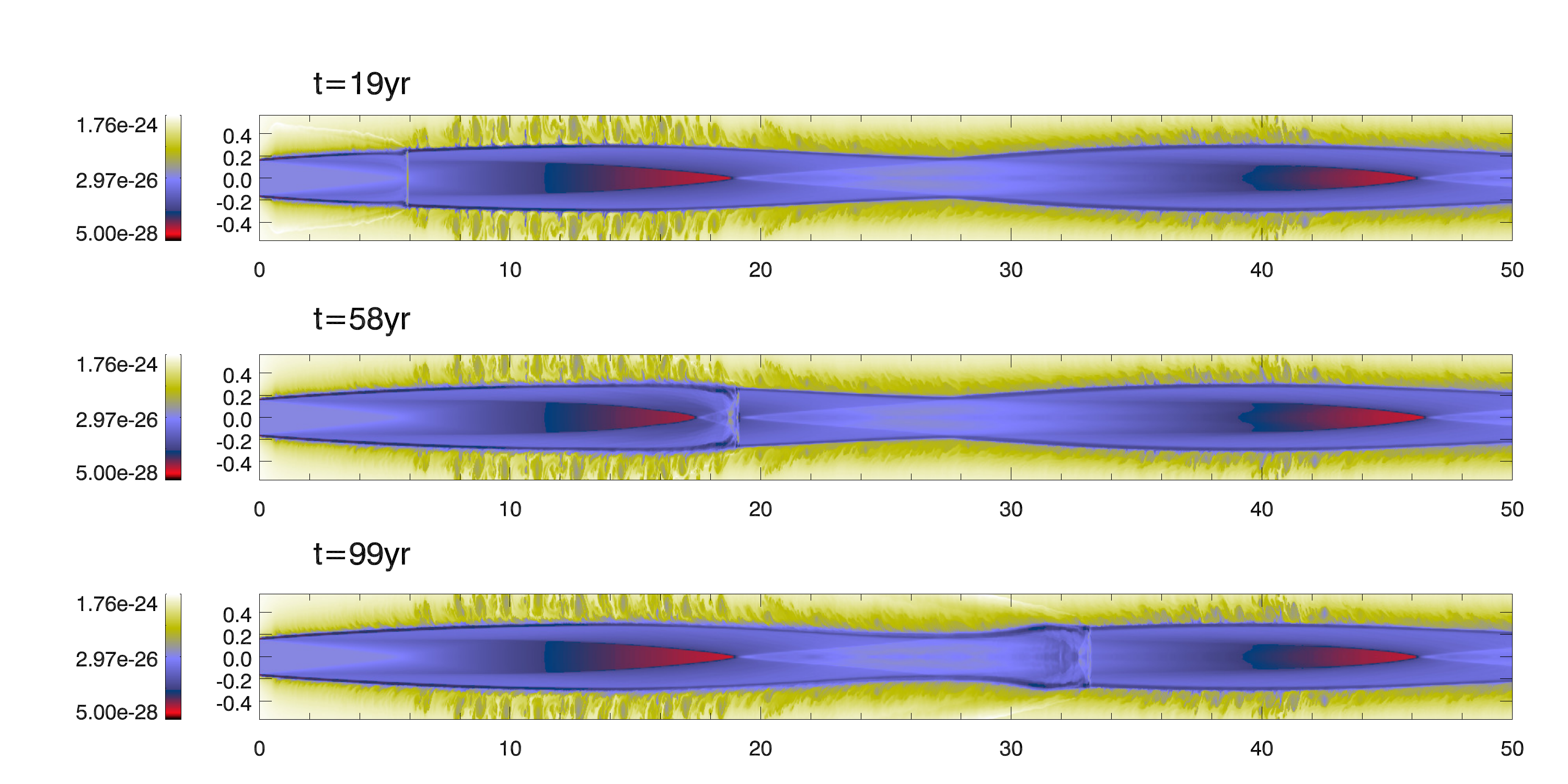}
\caption{Three snapshots of a two-dimensional, axisymmetric numerical simulation of a perturbed overpressured jet following the evolution of the perturbation (see text). Taken from \cite{fr13}.}
\label{fig-4b}       
\end{figure*}

  A numerical simulation of a perturbation injected in an overpressured jet with respect to the ambient medium was performed, and the emission at a viewing angle of 90$^\circ$ was computed. More complete emission simulations at small viewing angles are being performed to confirm this result (Fromm et al., in preparation). These simulations are similar to those presented in \cite{go97,ko97}, although now the interest lies on the spectral evolution of the perturbed jet, more than in the expected flux increase as obtained in the aforementioned references. The initial jet overpressure ratio is $P_j/P_a=3$, the jet Lorentz factor is $\gamma=12$, adiabatic exponent is $\Gamma=13/9$, and the jet density ration relative to the ambient $\rho_j/\rho_a=0.02$. The resolution used is 32 cells/$R_j$ in the two dimensions, with a grid size of $300~R_j\times10~R_j$. The jet produces two conical recollimation shocks within the grid. After equilibrium is reached, a perturbation is introduced with a duration that takes into account the injection time-scales of one month (based on the typical duration of X-ray flares, \cite{ma02}). Fig.~\ref{fig-4b} shows three snapshots of the perturbation traveling through the jet in comoving (host galaxy) time (as viewed at 90$^\circ$).

  The general conclusion that can be extracted from this work is that the perturbation injected at the end of 2005 interacted with a stationary, recollimation shock and went through, probably, an adiabatic compression that generated an increase in the flare flux. The position of this stationary feature does not coincide with the core position, but it is within the core region. The analysis of the core-position at the studied frequencies, up to 43~GHz, does not allow to know whether the core coincides with a still inner shock, but it seems to correspond to a $\tau=1$ surface, with $\tau$ the optical depth, up to that frequency \cite{fr13b}.  
  
\subsection{Implications}
\label{ssec-1-3}

  The initial hypothesis involving the interaction of a traveling perturbation in the jet and a recollimation shock has been confirmed up to our
resolution and technical limits. Although there could be other possible phenomena producing flares and changes in their spectral evolution,
we have collected enough evidence to claim that the most probable physical process in the case of the 2005 flare in CTA~102 is the ejection of a perturbation, in the form of a density increase, and the later interaction of this perturbation with a stationary shock. It has been shown that this latter process is able to produce an increase in the flux of the flaring region (i.e., the perturbation) and the observed spectral evolution. This fact could be taken into account as a signature of shock-shock interaction when studying the evolution of flaring events in blazars and correlations in a multi-frequency context. The global picture would thus be the following: A perturbation is injected into the jet, as an increase in the mass-flux in the jet. This perturbation generates a shock wave, compressing the flow and accelerating particles. In the most compact region, the particle and photon densities are large enough to produce intense inverse Compton scattering. As the perturbation expands downstream with the jet, the particle density drops, the region becomes optically thinner at more frequencies, and inverse Compton ceases to be the dominant loss channel. In the case of the studied flare, it seems that at this point, adiabatic losses are the dominant energy-loss mechanism. At some point, the perturbation enters into the recollimation region, which is not a discrete, planar region, but an extended conical one, in which the material in the perturbation gets gradually compressed up to the maximum compression at the tip of the cone formed by the standing shock. After the induced increase in flux, which is large enough to change the observed light-curve of the flaring event, the perturbation goes back into an adiabatic-loss stage. Farther downstream, similar events, although with much more modest increases in the flux can happen. This is due to the fact that the perturbation has already dissipated an important fraction of its kinetic energy and the capacity to generate remarkable increases in flux is lost.
  
\begin{figure}
\centering
\includegraphics[width=\columnwidth,clip]{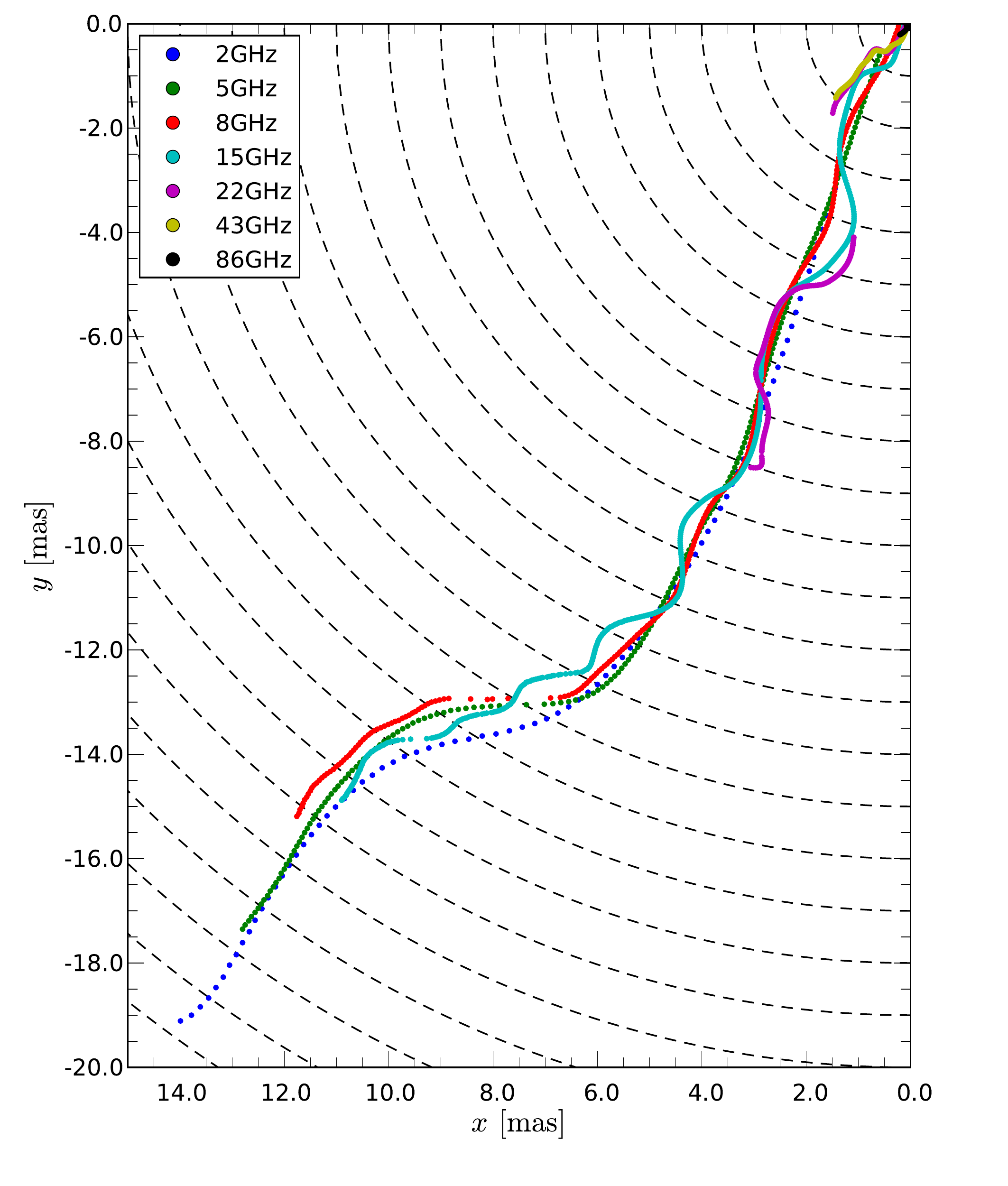}
\caption{Ridge-line of the jet in CTA~102 at different frequencies. Dashed lines indicate 1~mas increase in radial distance to the core. Taken from \cite{fr13b}.}
\label{fig-4a}       
\end{figure}

\begin{figure}
\centering
\includegraphics[width=\columnwidth,clip]{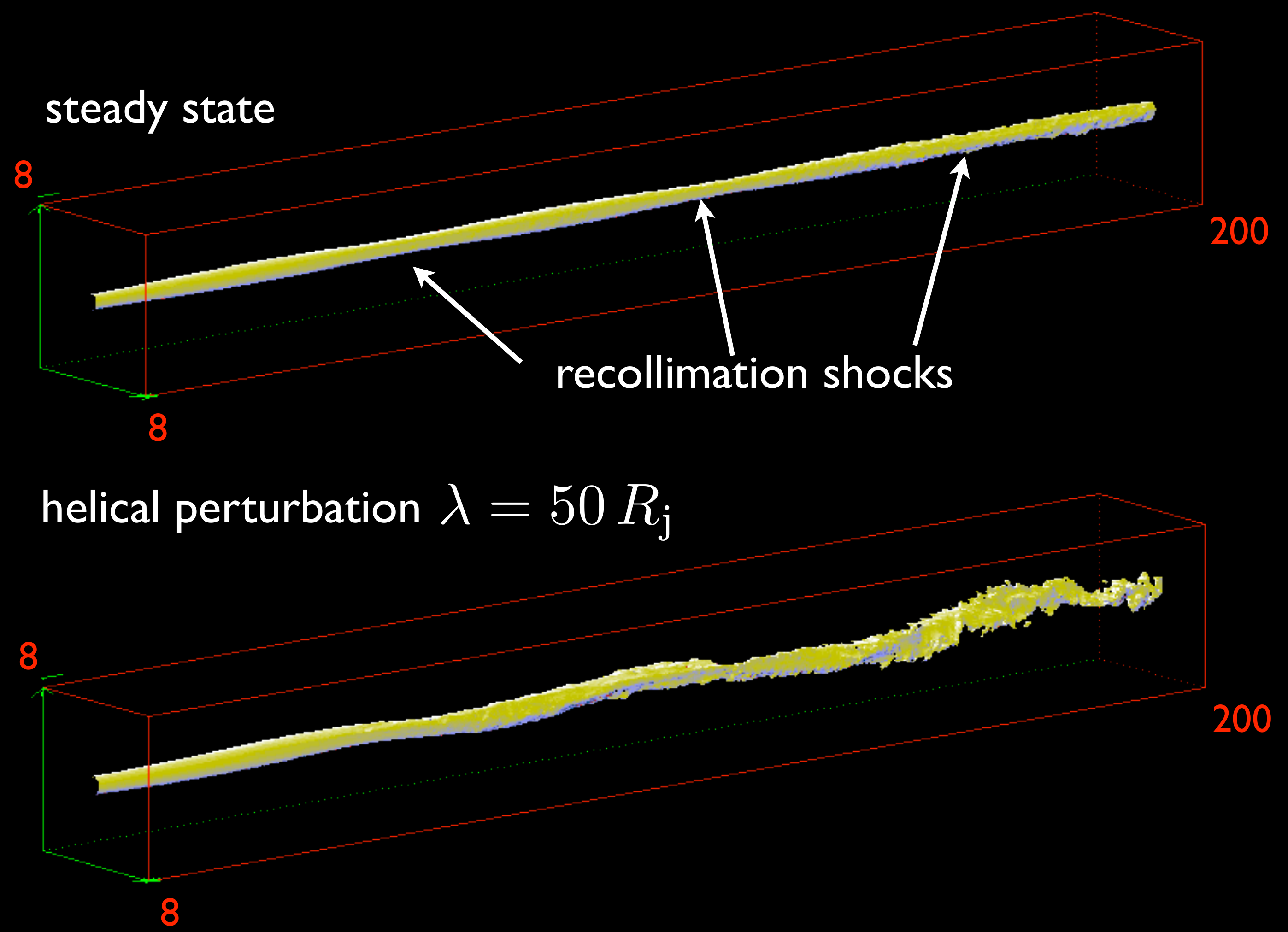}
\caption{Numerical simulation of an overpressured jet in which a helical perturbation is introduced. In the top panel, the overpressured jet in equilibrium is shown. In the bottom panel, the same jet after the helical perturbation has crossed the grid. The recollimation shocks and the helical instability compete to determine the jet morphology. Simulation performed with Ratpenat. Courtesy of C.M. Fromm.}
\label{fig-4c}       
\end{figure}

 The jet in CTA~102 shows a helical shape along distance (Fig.~\ref{fig-4a}, \cite{fr13b}). It could be (wrongly) claimed that a helical structure is incompatible with the presence of recollimation shocks. The fact that the development of these two different features is generated by independent causes (pressure mismatch between the jet and the ambient versus pressure mismatches within the jet) makes their existence in a jet completely compatible and possible, as a simple three-dimensional relativistic hydrodynamical simulation shows (see Fig.~\ref{fig-4c}, \cite{pe05}). The simulation shown in Fig.~\ref{fig-4c} consists of an overpressured jet with respect to the ambient that generates several recollimation shocks within the grid before reaching an equilibrium situation (top panel). The jet overpressure ration $P_j/P_a$ is 1.5, the jet Lorentz factor is $\gamma=5$, adiabatic exponent is $\Gamma=13/9$, and the jet density ration relative to the ambient $\rho_j/\rho_a=0.02$. The resolution used is 8 cells/$R_j$ in the three dimensions. After this equilibrium is obtained, a helical perturbation with a wavelength of 50~$R_j$ was introduced. The result is shown in the bottom panel, where a competition between the pinching and the helical structures can be observed. 

\section{Helical structures}
\label{sec-2}

\subsection{Evidence}
\label{ssec-2-1}

\begin{figure*}
\centering
\includegraphics[width=0.23\textwidth,clip]{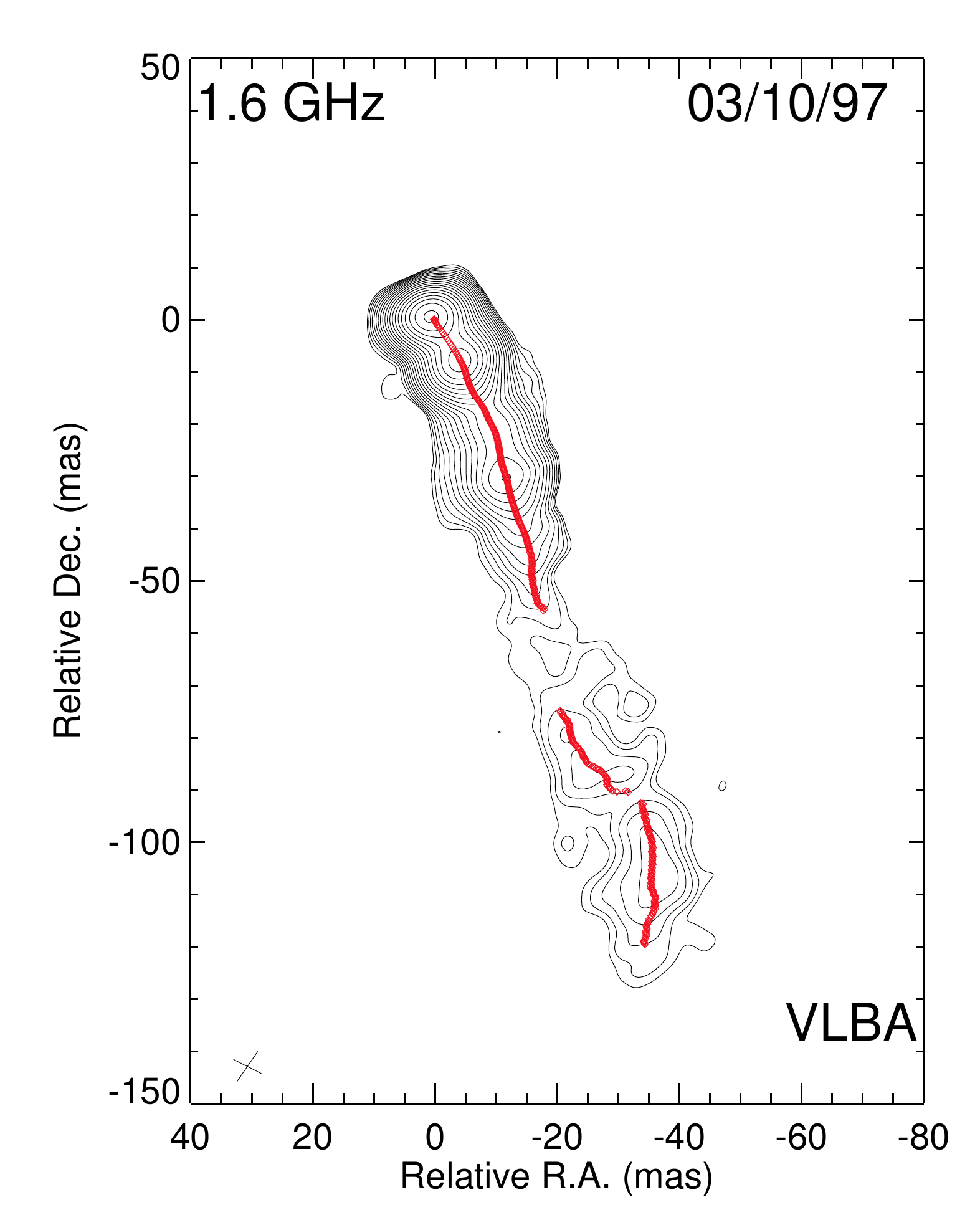}
\includegraphics[width=0.23\textwidth,clip]{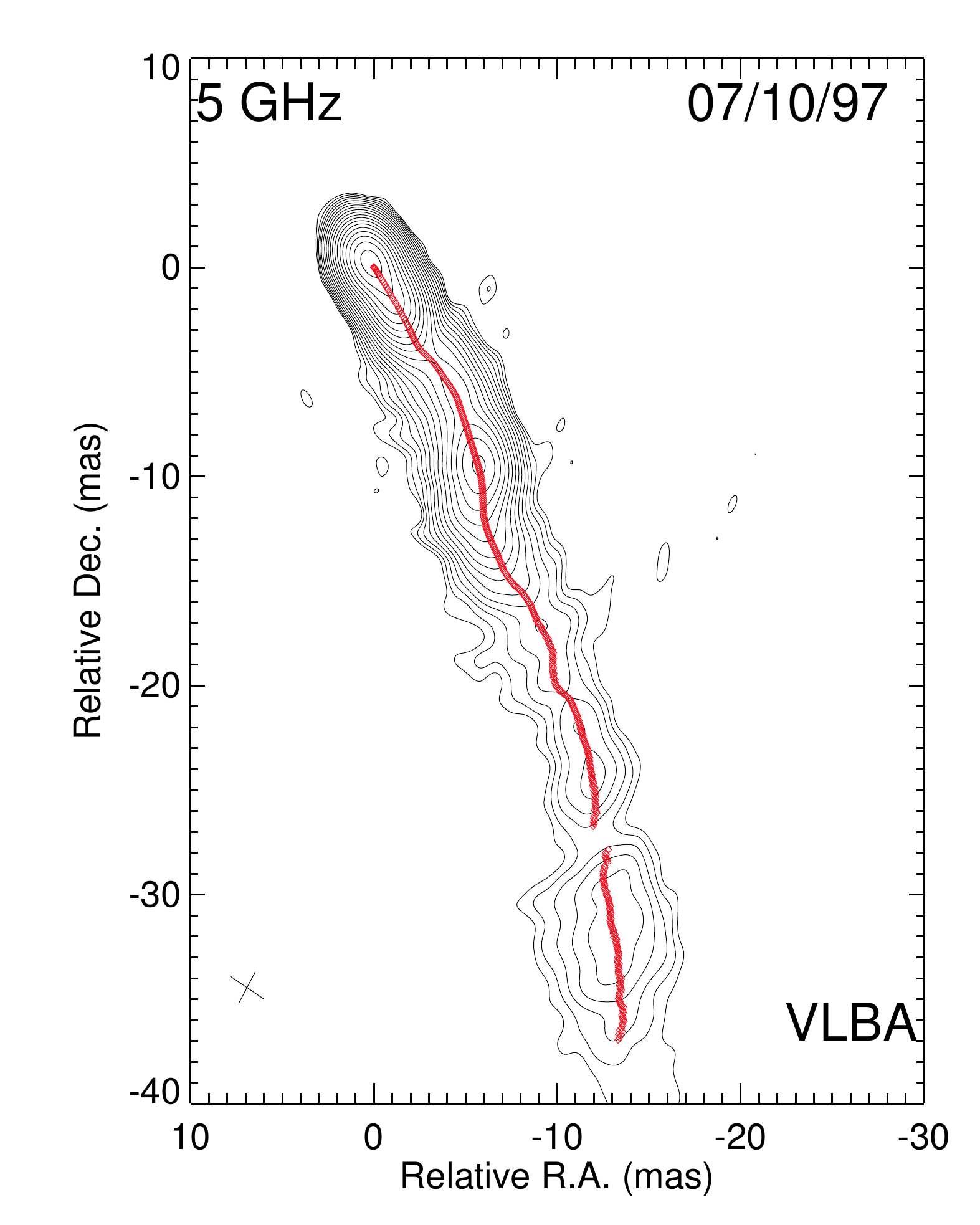}
\includegraphics[width=0.23\textwidth,clip]{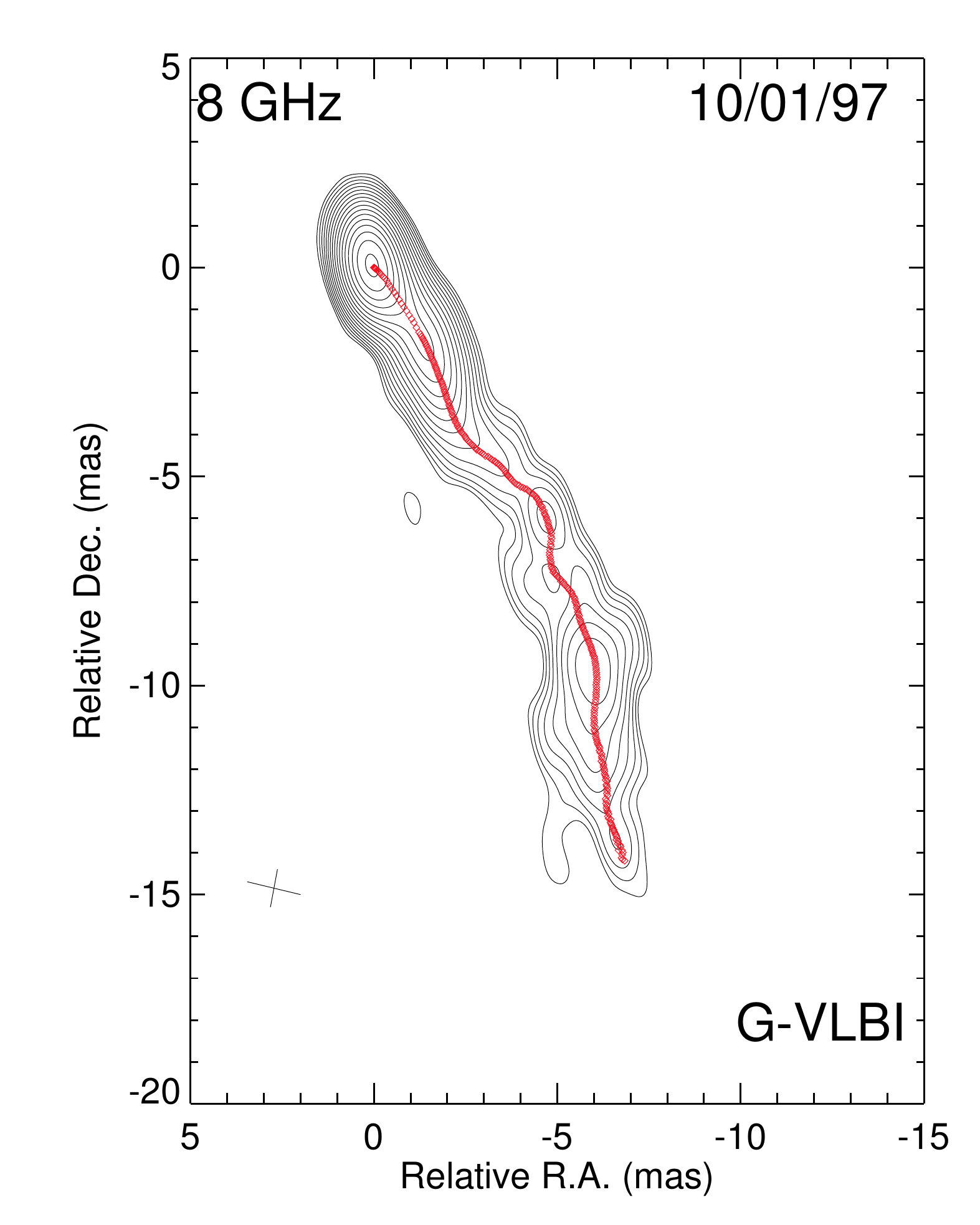}
\includegraphics[width=0.23\textwidth,clip]{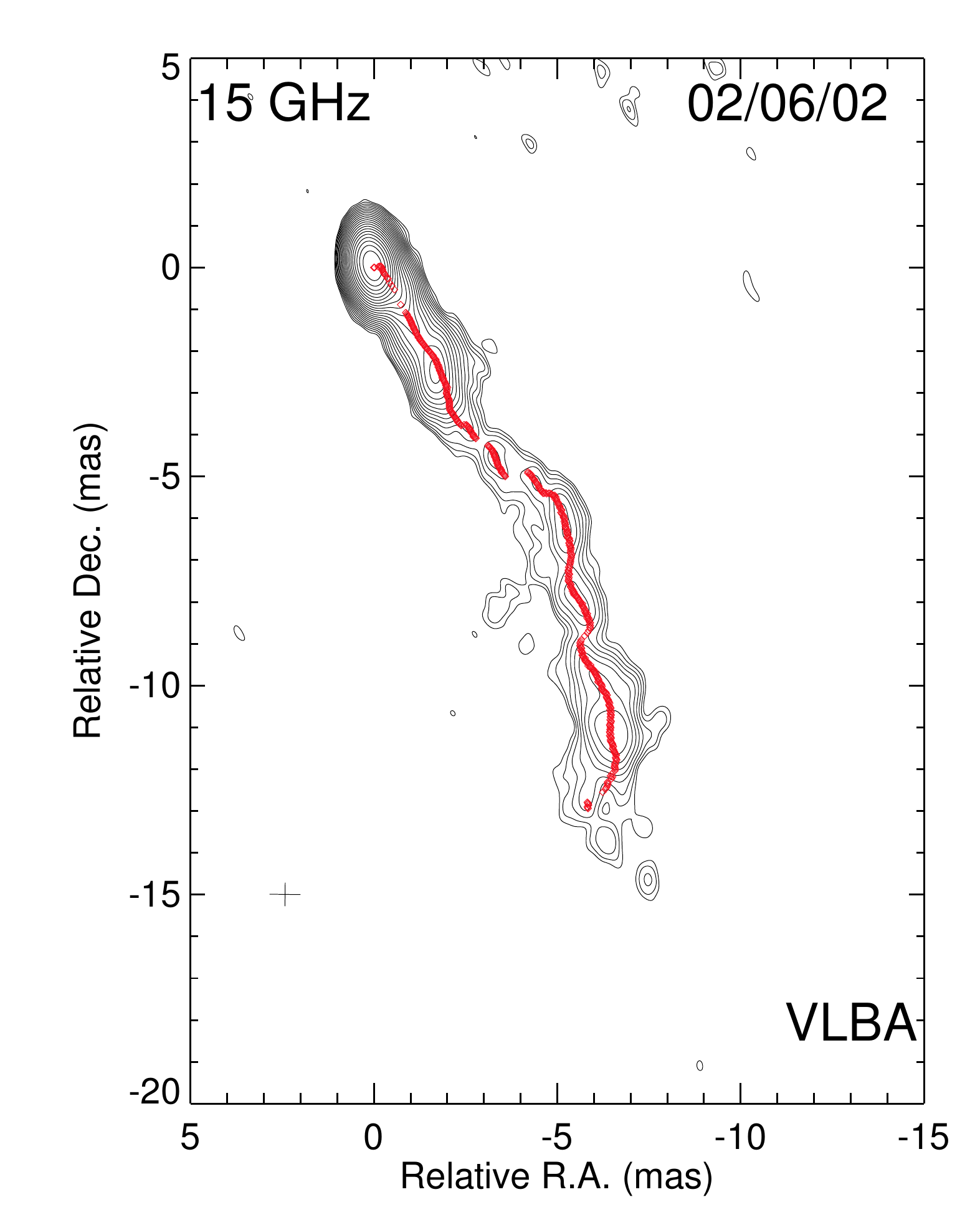}
\caption{The jet in 0836+710 at different frequencies and epochs. Notice the different scales. The ridge-line is overplotted. Taken from \cite{pe+12}.}
\label{fig-50}       
\end{figure*}

\begin{figure}
\centering
\includegraphics[width=\columnwidth,clip]{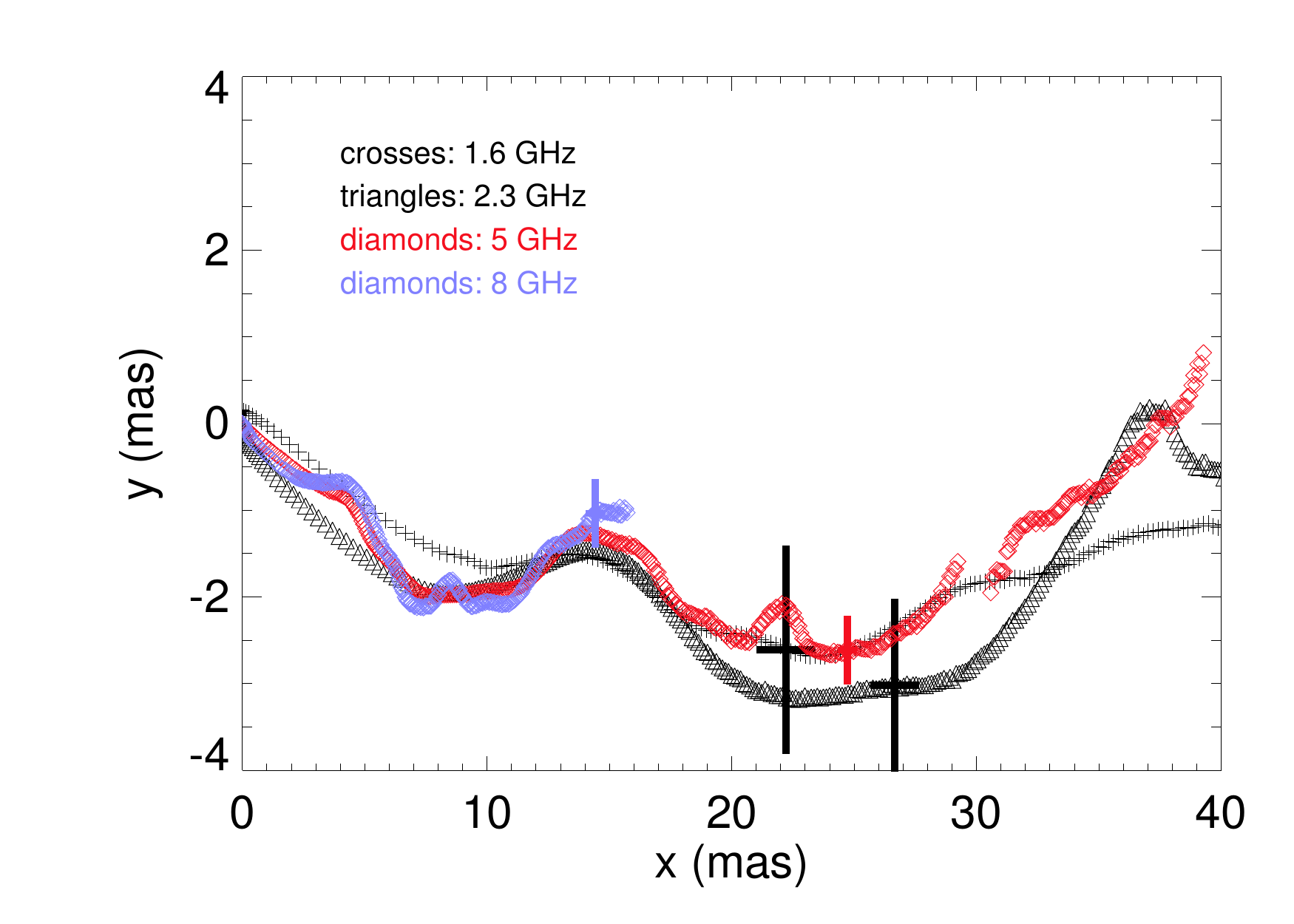}
\caption{Ridge-line positions along the (large-scale) jet propagation axis at different frequencies (1.6, 2.3, 5 and 8~GHz) in 1997. 
The crosses indicate estimates of errors in the determination of these positions (only indicated in one point per frequency for the sake of clarity).}
\label{fig-5}       
\end{figure}

\begin{figure*}
\centering
\includegraphics[width=\columnwidth,clip]{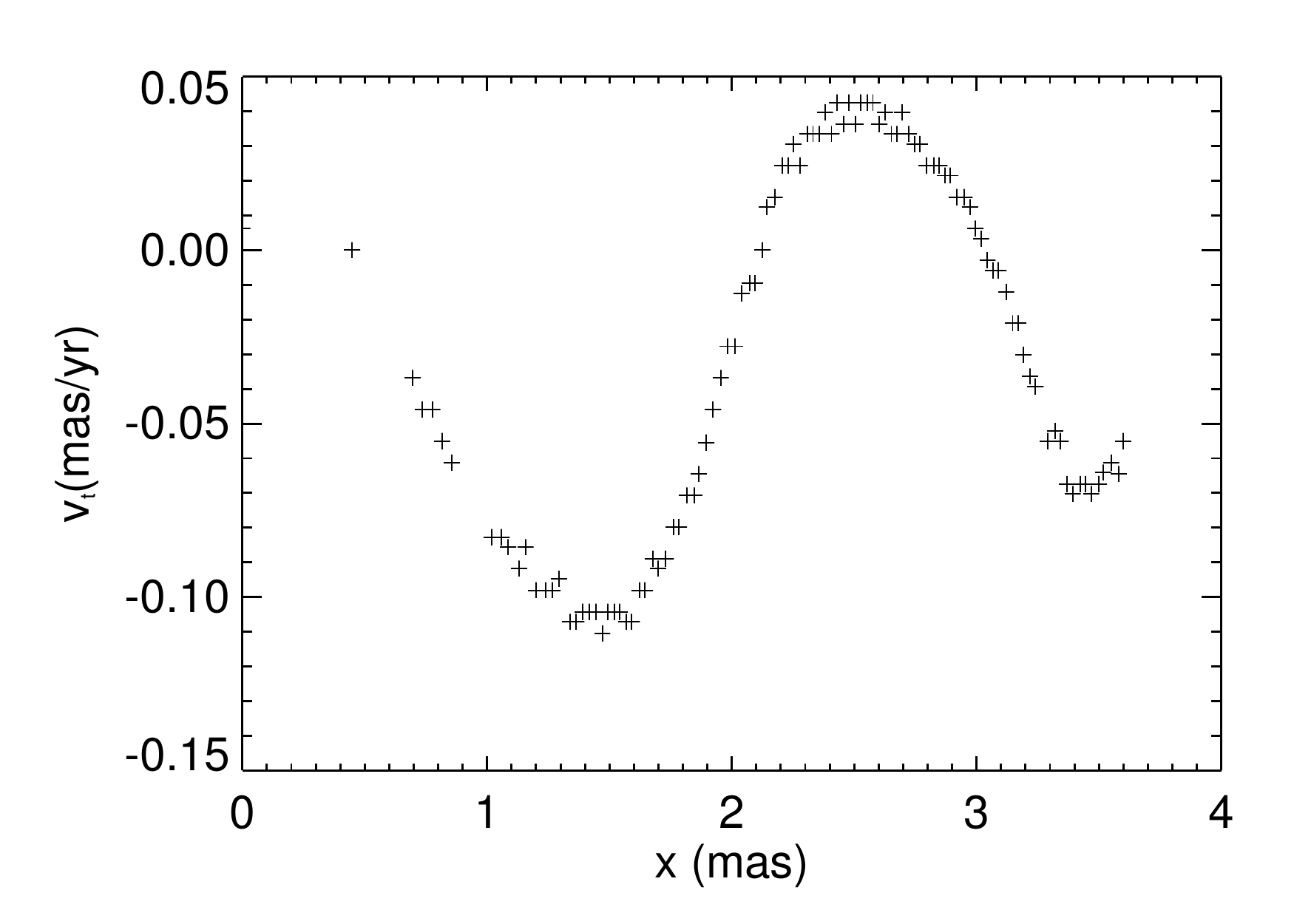}
\includegraphics[width=\columnwidth,clip]{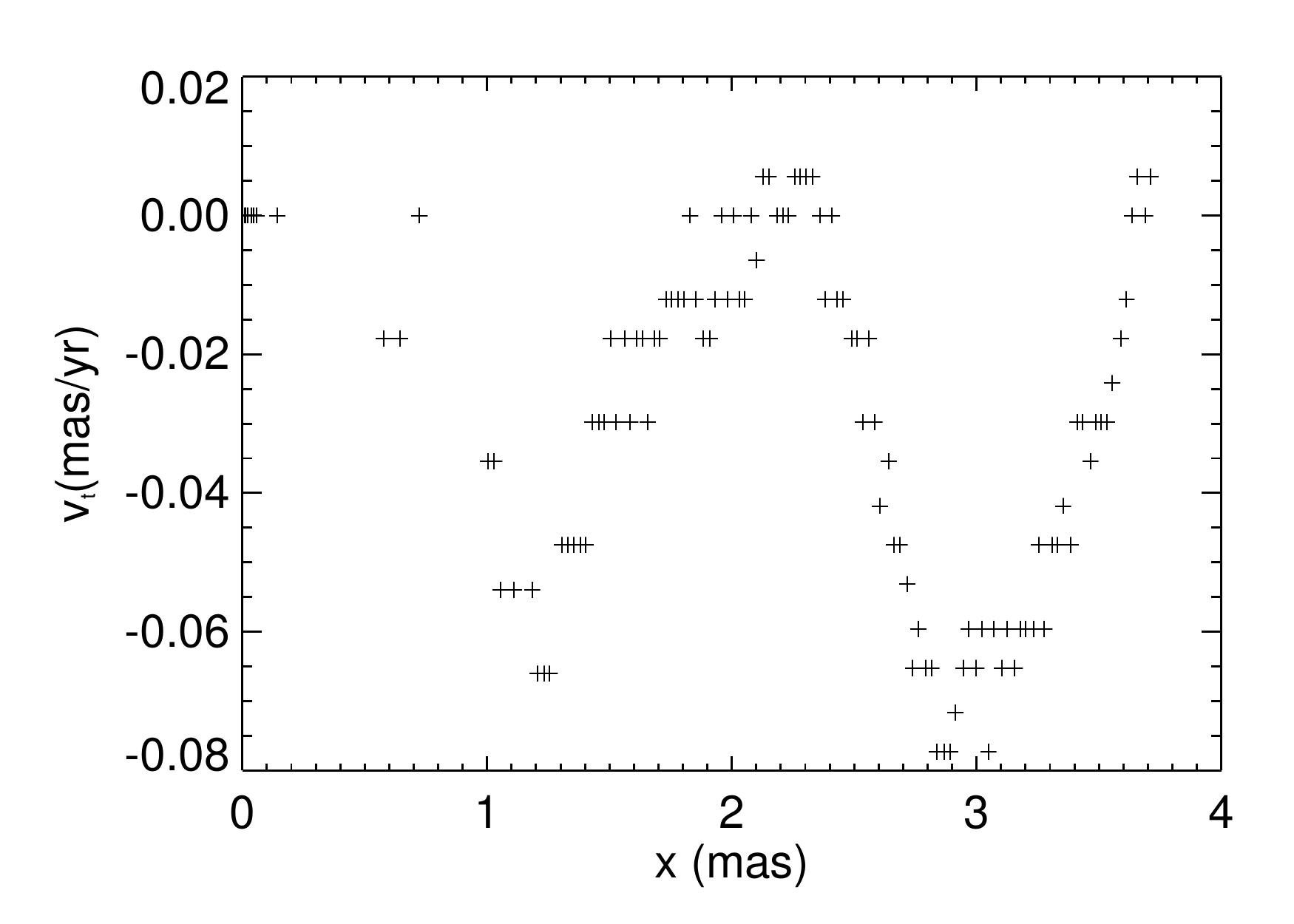}
\caption{Transversal velocity of the ridge line obtained from dividing the displacements of the ridge-line (at 15~GHz) between two pairs of epochs (2002-2003, left, and 2008-2009, right) by the time between them. Taken from \cite{pe+12}.}
\label{fig-6}       
\end{figure*}

\begin{figure*}
\centering
\includegraphics[width=\columnwidth,clip]{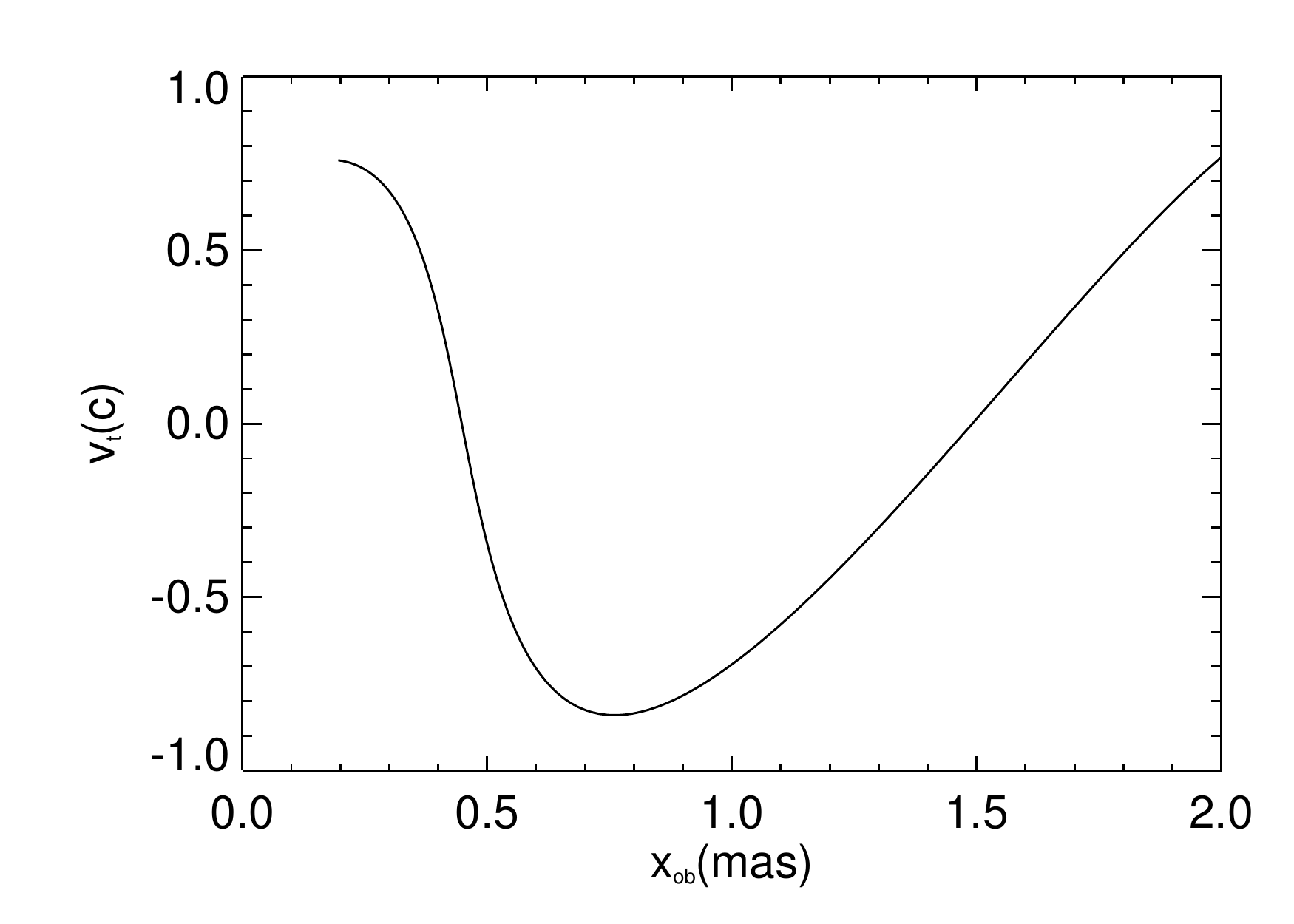}
\includegraphics[width=\columnwidth,clip]{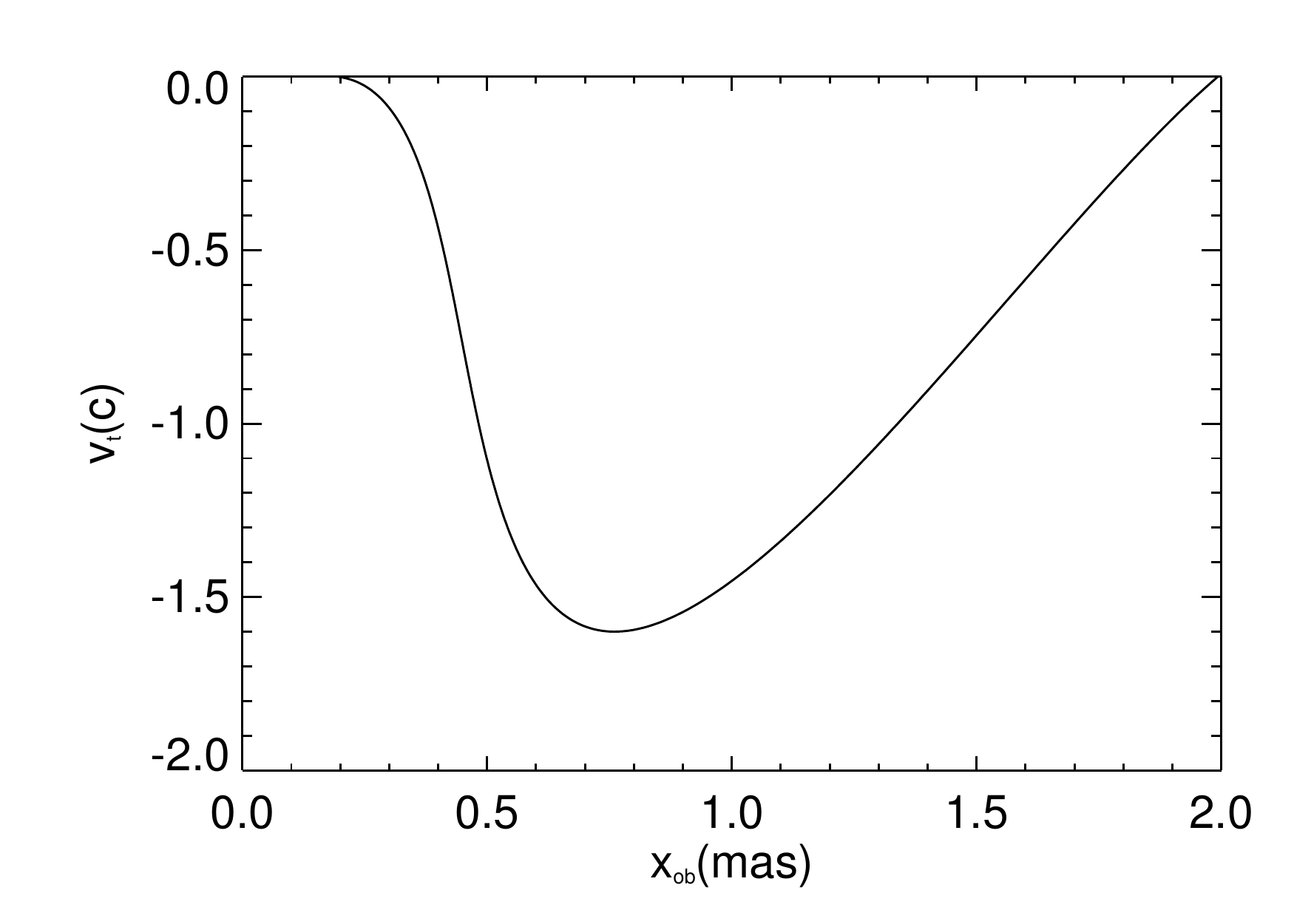}
\caption{Same as Fig.~\ref{fig-6}, obtained from an analytically generated three-dimensional helix. The left panel shows the result obtained by fixing the position of the core, whereas the right panel shows the result obtained by removing this assumption. By doing this, the superluminal transversal velocity is corrected in the latter. Taken from \cite{pe+12}.}
\label{fig-7}       
\end{figure*}

  A collection of observations at different frequencies of the jet in S5~0836+710 ($z=2.16$, $8.37$~pc/mas) was analyzed in \cite{pe+12} (see Fig.~\ref{fig-50}). The different observations of the jet at different frequencies ranging from 1.6~GHz to 43~GHz for a given epoch (or close enough observation epochs), give the same ridge-line positions within the errors. The ridge-line is defined as the peak of emission of a gaussian fitted to transversal cuts of the jet along the direction of propagation. Fig.~\ref{fig-5} shows the obtained ridge lines from 1.6 to 8~GHz, with estimates of the errors as one fifth of the beam size in the corresponding direction. This image serves as an example of those obtained for different frequencies and epochs and giving similar results (see also Fig~\ref{fig-4a} for the case of CTA~102). Some of the small wavelength oscillations could be attributed to uv-coverage effects, as discussed below, but the general trends are reproduced for all cases, and this can only be explained by the ridge-line of the jet corresponding to a region of maximum emission at all the observed frequencies, i.e.,
probably corresponds to the radiative signature of a physical structure, which is certified mainly at the frequencies at which the jet is transversally resolved: High resolution images at 15~GHz confirm that the ridge-line position does not necessarily coincide with the centre of emission of the radio jet \cite{pe+12}. Within the first milliarcsecond, these images at 15~GHz allowed the measurement of transversal displacements, which show a clear wave-like oscillation pattern with distance. Fig.~\ref{fig-6} shows the results for two pairs of observing epochs (2002-2003 and 2008-2009, separated by 0.82 and 0.75 years, respectively). The velocity patterns show similar shapes between them, albeit different amplitudes. The obtained ridge-line velocities at these scales are superluminal, but this is due, according to the authors, to: 1) relativistic wave velocities, 2) a small-scale oscillation of the core, very difficult to detect in this distant source, but observed in others (e.g., M~81, \cite{mv11}, see also Fig.~\ref{fig-7}), and 3) displacements of the ridge-line being smaller than the errors in the determination of its position. The latter aspect will be discussed in more depth in the next section. Finally, it is interesting to note that the amplitude of the oscillation, as seen at the lowest frequencies, grows along the jet propagation direction. This fact can be due to the coupling to an unstable mode of the Kelvin-Helmholtz or current-driven instabilities \cite{pe+12b}. 

  The evidence exposed in the previous paragraph fits with the suggested association of ridge-lines with maxima in emission produced by pressure enhancements within the jet, possibly coupled to the development of instabilities.

   Recently, Cohen et al. (this volume) have presented new results of the nearby source BL~Lac, using long-term monitoring of the source at 15~GHz, which confirm that the ridge-line of emission within the jet shows a wave-like behavior. In addition, it has already been reported, by using stacked images, that jets may show different structure with time, and fill a wider channel than observed at a single epoch, for instance in the case of 3C111 \cite{gro12}.
     
\subsection{Causes and implications}
\label{ssec-2-2}

  The origin of these waves is unclear because the observed patterns are not necessarily periodical, although there seems to be agreement in the possibility that the long-wavelength structures must be produced at the formation. The precession of this region, due to the accretion disk or a secondary black-hole orbiting the primary one has been invoked as a triggering mechanism. Shorter-term oscillations can be attributed to perturbations triggered by asymmetries in the jet itself or in the ambient medium, lateral winds or entrainment of clouds that rotate around the galactic centre from a side of the jet, for instance. However, it is difficult to disentangle the causes from the effects in these objects and this is more the case when the models proposed have a moderately large number of free parameters. In conclusion, different models can end up reproducing the observed patterns within the errors produced in their derivation. In the case of small wavelengths, some of them can even be observational artifacts, as discussed below. Another difficulty to be taken into account is the following: If the triggering process is continuous, the pattern can show a stable wavelength in distance and time. On the contrary, a single perturbation event can propagate downstream, coupled to an unstable mode, but it will only produce an isolated wave-pulse.
  
  A helical structure generated by the pressure maxima of a wave developing within a jet could be explained in an idealized way, by the following equations: 
  \begin{eqnarray}\label{hel}
  x_h =
A_0\,e^{z_h/\lambda_i}\cos\left(\frac{2\,\pi(z_h-v_w\,t)}{\lambda}
+\varphi_0\right) \nonumber \\
  y_h =
A_0\,e^{z_h/\lambda_i}\sin\left(\frac{2\,\pi(z_h-v_w\,t)}{\lambda}
+\varphi_0\right)\, ,
\end{eqnarray}

with $z_h$ the axis along which the helix is defined, $x_h$ and $y_h$ the transversal directions, $A_0$ the initial amplitude, $\lambda_i$ the growth length of the unstable mode, $\lambda$ the wavelength, $v_w$ the wave velocity, which is independent of the flow velocity, and $\varphi$ the initial phase. The relation between the observed ($x_{ob},y_{ob}$) coordinates on the plane of the sky and the intrinsic ($x_h,y_h,z_h$) ones, as specified above, is the
following, accounting for relativistic and projection effects:
 \begin{eqnarray}\label{proj}
x_{ob} = x_h(z_h) \left[ \cos \alpha - \left( {\sin^2 \alpha \over 1 - \beta_w
\cos \alpha } \right) 
\beta_w \right]+ \nonumber \\ 
z_h {\sin \alpha \over (1 - \beta_w \cos \alpha)} \nonumber \\
  y_{ob} = y_h\, ,
\end{eqnarray}
where $y_ob$ is selected as the direction perpendicular to the jet axis and parallel to the plane of the sky. These equations, in the absence of motion, reduce to:
 \begin{eqnarray}\label{proj2}
  x_{ob} = x_h\cos \alpha + z_h\sin \alpha \nonumber \\
  y_{ob} = y_h.
\end{eqnarray}
  If there are more waves evolving, the picture gets more difficult to interpret, as each wave may have a different propagation velocity and wavelength.      
  
 There is open debate as to whether these waves are magnetic or purely hydrodynamical in nature. The claim that the waves correspond to magnetic instabilities is based on the fact that the observed radio components seem to follow the ridge-line as they propagate (Cohen et al., this volume, and \cite{na10} for the interpretation of zig-zag motions in jets): If the jet ridge line is identified with a kink instability in the magnetic field of a magnetically dominated jet, the flow would be forced to follow this kink. 
  
   This argumentation follows from the idea that if the jet is strongly magnetized, the flow always follows the field lines and thus shows a twisted path if the field is helically twisted, whereas in the kinetically dominated case the flow could follow a straight stream-line even if the radio-jet shows a helical structure, because 1) the helical structure responds to the bright high pressure region within a wide channel, or 2) due to ballistic motion with changing injection direction, as seen in the microquasar SS~433 (e.g., \cite{st02}). However, in the case of a growing instability the flow can propagate along straight stream-lines while the amplitude of the instability is not large. When it grows to nonlinear values, the jet flow starts to deviate from its original trajectory by the pressure gradients within the jet (see, e.g., \cite{mi11,pe05,pe06} for magnetized and non magnetized jets). It was shown via numerical simulations \cite{pe06} that an asymmetric perturbation at the base of a relativistic jet propagates downstream and naturally creates a pressure maximum helical structure, which can eventually distort the jet flow and force it into a helical path when its amplitude grows to large enough values. Within the same simulation setup, perturbations in flow density were injected into the jet. The perturbations generate shock waves, which propagate ballistically along the first part of the simulated jet, but end up following the helical path when most of their kinetic energy has been dissipated. It has also been been reported that the development of the helical Kelvin-Helmholtz instability can force the jet flow into a helical path, albeit different from the helicity of the pressure maximum, with the helicity of this maximum showing larger amplitude than that of the jet flow \cite{ha00,ha03}. The larger the Lorentz factor of the flow and the shorter the wavelength of the mode, the more different is the helicity of the flow as compared to that of the pressure maximum. This can be understood in terms of the larger inertia of the jet flow with increasing Lorentz factor \cite{ha00,pe05}. 
   
   
   Finally, let us stress that the argument of the flow following the field lines also works in the opposite way: a particle dominated plasma with an ordered velocity field would force the magnetic lines into the same path. Thus, a purely hydrodynamical jet could produce observational patterns similar to a strongly magnetized jet, and further evidence has to be given in any of both directions to make clear statements on the issue.    

    Moreover, the observed helical ridge-line structure could be an observational bias produced by differential emission within the jet: If the obtained ridge-lines correspond to the pressure maximum produced by a kink current-driven instability (see, e.g., \cite{mi11}) or helical Kelvin-Helmholtz instability (see, e.g., \cite{pe06}), a high-pressure region and a low-pressure one can be found within the jet cross-section. In this case, the passage of a ballistic component following a straight path along the jet, could enhance the emission of the whole jet, but the effect would be more visible on the ridge-line (high-pressure region). As a result, any radio-component fitted to the region would be centered there. Then, if the ridge-line corresponds to a propagating wave, then the fitted-components would also move apparently along the jet and, of course, on top of the ridge-line. It is thus crucial to disentangle the nature of the components and whether or not they follow the ridge-line. 

     The waves that were associated with the ridge-line could be coupled to a growing instability, as indicated by the increasing amplitude of the helix with distance in the case of S5~0836+710, as seen at 1.6~GHz. If so, it is difficult to distinguish between current-driven instability and Kelvin-Helmholtz instability modes, both being solutions to the linearized relativistic, magnetized flow equations and both being possible sources of helical patterns \cite{ha07}. Recent work on current-driven instability \cite{mi11} shows that a helical kink propagates with the jet flow if the velocity shear surface is outside the characteristic radius of the magnetic field, i.e., approximately the radius at which the toroidal magnetic field is a maximum. If the observed pattern corresponds to a current-driven kink instability, the observed transversal oscillation in the jet of 0836+710 requires that the kink be moving with the flow and implies that the transversal velocity profile is broader than the magnetic field profile, i.e., the velocity shear surface lies outside the characteristic radius of the magnetic field. Thus, in this case the jet would have a magnetized spine surrounded by a particle dominated outer region and a current-driven kink in the spine would move with the flow. However, this cannot be confirmed by present data.

\subsection{Caveats}
\label{ssec-2-3}
 
\begin{figure*}
\centering
\includegraphics[width=\columnwidth,clip]{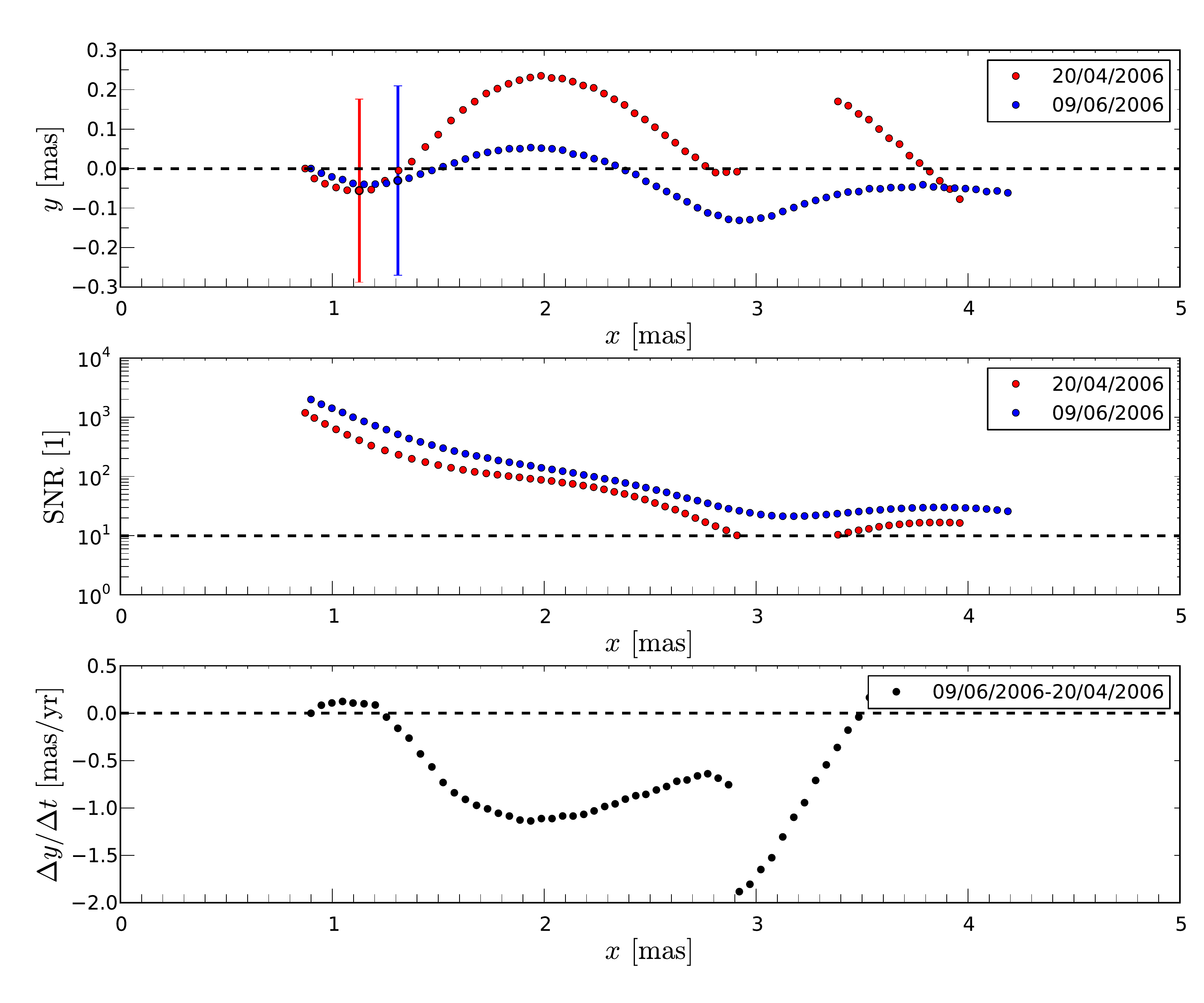}
\includegraphics[width=\columnwidth,clip]{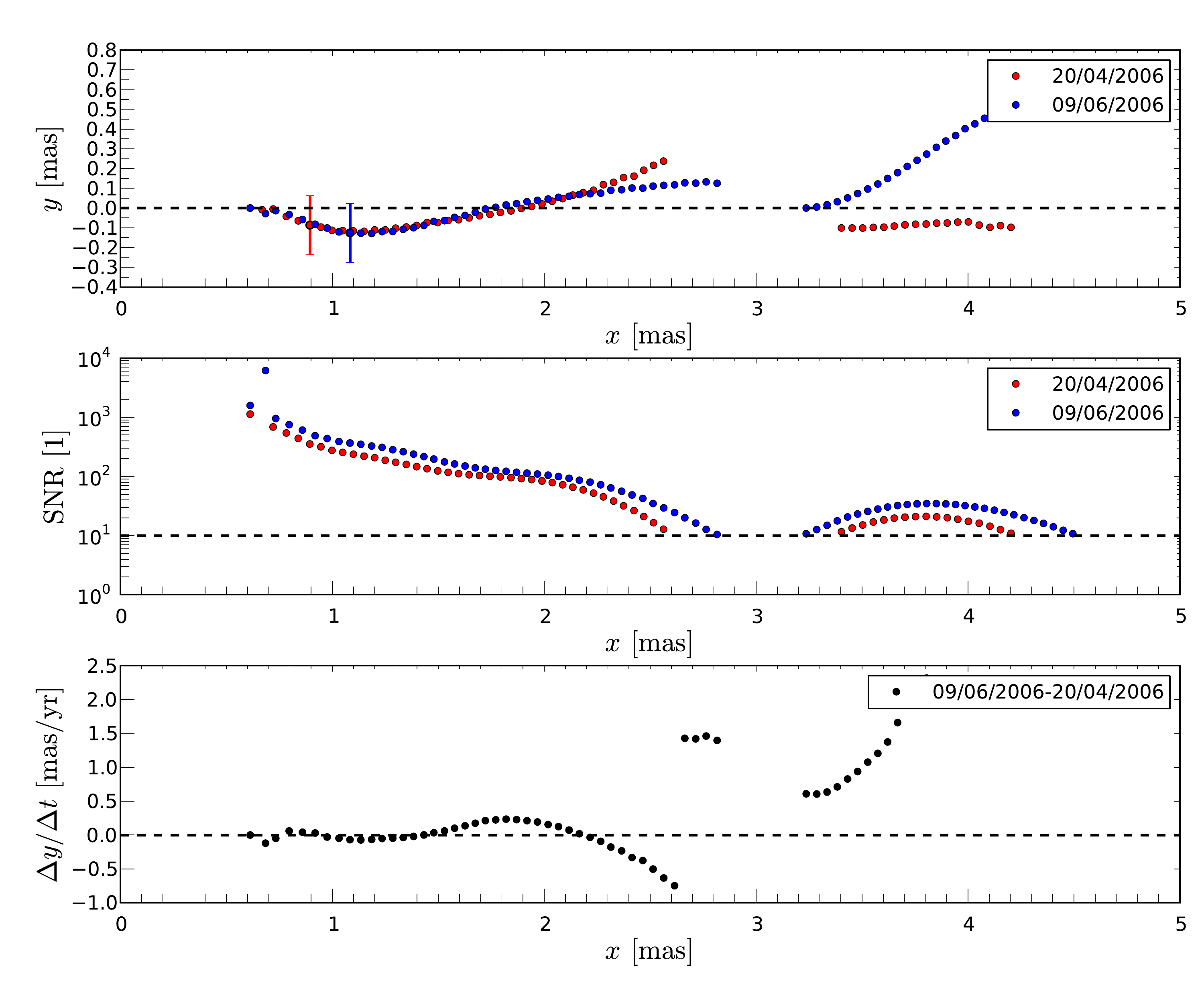}
\caption{Jet ridges of an artificial jet at two different epochs (top panels), signal-to-noise ratios (central panels) and transversal displacements (bottom panels), as obtained with an elliptic beam (left column) and circular beam (right column). The error bars apply to all the points and are taken as one fifth of the maximum beam size. Fromm et al. (in preparation).}
\label{fig-9}       
\end{figure*}

   The detection of a wave pattern in the oscillation of the ridge-line between different epochs has been invoked as a proof of the real existence of waves in relativistic jets (see Section~\ref{ssec-2-1} in this work, \cite{pe+12}). However, the amplitude of the oscillations is smaller than the errors in the determination of the position of the ridges. Therefore, this proof has to be taken with caution and reviewed as possibly generated as an observational artifact. With this aim, possible changes in the ridge line have been studied, by deriving it in the case of an artificially generated straight jet at different epochs (and different uv-coverage, Fromm et al., in preparation). The jet is produced with a gaussian transversal structure centered on its axis. The ridge line should be straight, and this is the case, but for irregular oscillations within errors, when one of the axis of the deconvolution beam of the VLBI observation coincides with the jet axis (or if the beam is circular). On the contrary, if there is an angle between the beam axis and the jet axis, an artificial, regular oscillation of the ridge-line may be produced, due to the brightness distribution induced by the shape of the beam. Figure~\ref{fig-9} shows the results obtained for an elliptic beam with an angle between its axis and the jet axis (left column), and a circular beam (right column), applied to the same straight jet. The top panels show the location of the ridge along the jet axis, the central panels show the signal-to-noise ratio (SNR), and the last shows the velocity of the displacement between the epochs, in mas/yr (let us remark that the conversion at $z\simeq2$ between angular size and distance is $\simeq 8.4$~pc/mas). The bottom panel in the left column shows that wave-like oscillations can be artificially induced by applying an elliptic beam whose axis have an angle to the jet axis. From the upper and central panels in the right column, corresponding to a circular beam, shows that the deviation from a straight line can also be exaggerated at low values of the SNR (below 10). Although in the case of the circular beam a deviation is obtained, it has an irregular aspect without any hint of wave-like structure. The differences between epochs are mainly produced by the different distribution of the antennas in the uv-plane during the year. 
   
        In the case of the results given in Fig.~\ref{fig-6}, the beams at the 2008 and 2009 epochs are almost circular, whereas they are more elliptic in the case of the epochs in 2002 and 2003\footnote{see the images of the jet S5~0836+710 in the MOJAVE database \cite{li09}: http://www.physics.purdue.edu/astro/MOJAVE/sourcepages/0836+710.shtml}. Finally, it should be noted that, in order to obtain the transversal displacements represented there, it was assumed that $x_{ob}(t_1)=x_{ob}(t_2)$, without taking into account possible changes due to the time dependence given by Equation~\ref{proj}. Despite these facts, it is remarkable that the maxima and minima are obtained at similar positions for the two pairs of epochs, even though the four observations were done at different times of the year and that the time intervals between the two pairs of epochs are also different (0.82 and 0.75 years, respectively). Thus, a full statistical study should be performed in order to understand how probable it is to obtain regular structures from different uv-coverages between epochs, and that this regular structure preserves its global structure with time before being fully convinced that the wave motions are real and it can be used as an evidence of the presence of waves in jets. The rest of the evidence provided is not subject to the same caveat and therefore remains valid.

 \section{Summary and Discussion}
\label{sec-3}  
\subsection{Summary}
    As a result of a flaring event by the end of 2005, the spectral evolution of the source at GHz frequencies shows a pattern that can be identified with the shock-in-jet model. An unexpected change in the spectral evolution of the perturbation cannot be attached to a change in the viewing angle towards the traveling perturbation or to any new flaring event and an interaction with a standing, recollimation shock is suggested as a possible explanation. Compelling evidence is found for the presence of recollimation shocks in the jet of CTA~102. Numerical simulations confirm that shock-shock interaction can produce flux increase during a flaring event and can thus be responsible for changes in the expected spectral evolution. As a conclusion, it is claimed that the change in the spectral evolution of the flare is produced by the adiabatic compression of the perturbation as it interacts with a recollimation shock located at $r\simeq0.1$~mas from the core.
    
    It is possible to associate the ridge-line of emission of helical jets to wave patterns tracing a (gas or magnetic) pressure maximum produced by the growth of instabilities. It is however difficult to extract information on the nature of the waves due to relativistic and projection effects plus the errors made in the calculation of ridge-line position due to observational limitations. The core of the jet could also be oscillating and this motion should be tested and corrected in order to study the parsec-scale kinematics of the radio-components in jets. Using a non-steady point as a reference position for measuring motions in jets with very small viewing angles at moderate or large redshifts may bring to wrong conclusions regarding the transversal velocities of the radio-components, as could be the case in sources like NRAO~150 or OJ~287 \cite{ag07,ag12}. Small-scale oscillations of the ridge-line between epochs could be artificially induced by restoring beams forming an angle with the jet axis.
      
   Further combined observational and theoretical studies like this one are required to get more information on the nature of the growing instability and on the properties of the jet flow. High-resolution observations with the space-telescope \emph{RadioAstron} plus ground-based VLBI are programmed and could favor improved measures of ridge-line positions to be compared with previous ones.

\subsection{Implications for jet stability and dynamics}

  Both recollimation shocks and helical instabilities can have an important role in the long-term stability and dynamics of extragalactic jets. The former depend on pressure differences between the jet and the ambient, and can be enhanced by a drop in the ambient density. The jet ambient can vary significantly with time: When the jet is generated and during the first $10^5-10^7$~yr, depending on the jet power and ambient properties, the bow-shock/backflow (cocoon) determine the jet environment. The sound speed of the gas in the cocoon is high, so the pressure is basically constant through it. During this time, as the jet expands, the cocoon grows and its pressure drops. If the conditions at injection remain fairly constant, this situation can, on the one hand, produce jet overpressure and recollimation shocks. On the other hand, if the jet is underpressured with respect to the cocoon, during the initial expansion phases, an initial collimation shock is followed by expansions and recollimations, too. When the bow-shock is far out of the host galaxy and the jet has an age $>>10^7$~yr, the cocoon disappears after mixing with shocked interstellar gas and expanding, and is forced into something similar to the original galactic gas profile, which is determined by the gravitational potential. In this case, the jet flow might become overpressured as it propagates through the galaxy and the ambient gas density and pressure fall. Again, recollimation shocks can be generated. The factors that determine the location of these shocks are the jet overpressure ratio with respect to the ambient, the jet Lorentz factor and the pressure profile of the ambient \cite{dm88,fa91}. 
  
   At these shocks, kinetic energy is dissipated into internal energy and adiabatic compression, or particle acceleration can occur. Numerical simulations show that small jet overpressure (of factors 2-3) are enough to produce strong shocks and enhancements of emission and observed spectral evolution (Fromm et al., in preparation). If the opening angle of the jet is large, i.e., if the overpressure ratio is large, a Mach disk can be formed. This planar shock decelerates the flow to subsonic velocities and, even though reacceleration of the flow may occur in the following expansion, the fate of the jet is determined by mixing and deceleration. This is possibly the case of FRI jets in the so-called \emph{flaring region} \cite{pm07}. Nevertheless, in this proceedings I have focused on conical shocks produced in the case of small (but still larger than 1) overpressure ratios jets and their observational signatures.
    
  
   The development of helical instabilities can also play an important role in the long-term stability of jets. It has been claimed \cite{pe+12b} that the jet
in 0836+710 is possibly disrupted by the helical mode that produces the bends in the radio-jet at tens of mas from the core. The growth of the instability dissipates jet kinetic energy into thermal energy and produces jet deceleration. In addition, if the amplitude of the wave grows to nonlinear values, it can produced rapid jet disruption and significant deceleration to sub-relativistic velocity via mixing with the ambient medium \cite{pe05}.

%

%
%

\section*{Acknowledgements} 
I acknowledge C.M. Fromm for providing figures and interesting discussions, and for his work during the last years. I acknowledge P.E. Hardee, Y.Y. Kovalev, A.P. Lobanov, E. Ros, T. Savolainen and I. Agudo for their work within the projects summarized in these proceedings. This research has made use of data from the MOJAVE database that is maintained by the MOJAVE team \cite{li09}. I acknowledge financial support by the Spanish ``Ministerio de Ciencia e Innovaci\'on''
(MICINN) grants AYA2010-21322-C03-01 and AYA2010-21097-C03-01.

\end{document}